\documentclass[fleqn,usenatbib]{mnras}

\usepackage{newtxtext,newtxmath}
\usepackage[T1]{fontenc}

\DeclareRobustCommand{\VAN}[3]{#2}
\let\VANthebibliography\thebibliography
\def\thebibliography{\DeclareRobustCommand{\VAN}[3]{##3}\VANthebibliography}

\usepackage{graphicx}	% Including figure files
\usepackage{amsmath}	% Advanced maths commands

\usepackage{booktabs} % Timmi, for tables
\newcommand{\kms}{\,km\,s$^{-1}$} % kilometres per second
\newcommand{\kmso}{\,km\,s$^{-1}$\,} % kilometres per second
\newcommand{\mt}[1]{\mathrm{#1}}

\newcommand{\Mo}{$\mathrm{M}_{\odot}$\,}
\newcommand{\MO}{$\mathrm{M}_{\odot}$}
\newcommand\Tstrut{\rule{0pt}{2.6ex}}         % = `top' strut
\newcommand\Bstrut{\rule[-1.5ex]{0pt}{0pt}}   % = `bottom' strut
\defcitealias{Messa2018}{M18a}
\defcitealias{Messa2018b}{M18b}
\title[The impact of GMCs on clusters in M51]{The young stellar clusters in M51 and the impact of GMC encounters}

\author[T. G. J\o rgensen et al.]{
Timmi G. J\o rgensen,$^{1}$\thanks{E-mail: timmi.jorgensen@fysik.lu.se}
Ross P. Church,$^{1}$
\\
$^{1}$Lund Observatory, Division of Astrophysics, Department of Physics, Lund University, Box 118, SE-22 100 Lund, Sweden\\
}

\date{Accepted XXX. Received YYY; in original form ZZZ}

\pubyear{2025}

\begin{document}
\label{firstpage}
\pagerange{\pageref{firstpage}--\pageref{lastpage}}
\maketitle

% Abstract of the paper
\begin{abstract}
We investigate the young stellar cluster population of M51 and how it is affected by encounters with giant molecular clouds (GMCs). We combine a galactic model with $N$-body simulations of 5000 unique clusters in the mass range $[600-24000]$ \MO. We simulate each cluster twice: with (C$_{\rm{R}}$) and without (C$_{\rm{N}}$) tidal perturbations from the GMCs. We are able to reproduce the majority of the observed mass- and age functions. However, for the age function we see a large discrepancy for clusters with masses $\sim$ 5000 \MO, which is likely related to incompleteness in the observations. We find that old low-mass clusters, located close to the galactic centre, are most likely to be disrupted. The effect of GMC encounters causes a decrease in survivability by 8 per cent points for the oldest clusters with initial masses below 6000 \MO. For 15 clusters we find that the presence of GMCs can protect the C$_{\rm{R}}$ clusters from the rest of the galactic tidal field and thereby cause them to retain a significantly higher fraction of stars compared to the C$_{\rm{N}}$ clusters. For clusters that are super-virialized we find that the C$_{\rm{R}}$ clusters have a higher virial ratio compared to the C$_{\rm{N}}$ clusters, which is a result of interactions with GMCs. We see no significant difference between the C$_{\rm{R}}$ and C$_{\rm{N}}$ populations, indicating that over a time period of 200 Myr the effect of the GMCs cannot be detected. 
\end{abstract}

% Select between one and six entries from the list of approved keywords.
% Don't make up new ones.

%%%%%%%%%%%%%%%%%%%%%%%%%%%%%%%%%%%%%%%%%%%%%%%%%%

%%%%%%%%%%%%%%%%% BODY OF PAPER %%%%%%%%%%%%%%%%%%

\section{Introduction}
The lack of old open stellar clusters was first noted by \citet{Oort1958} who concluded that there must be an restriction on how old a cluster can become. \citet{Wielen1971} later showed that there is a lack of open clusters older than a few Gyr in the solar neighbourhood. A cluster's evolution is set by the balance between internal and external effects. The internal effects of stellar evolution \citep{Lamers2010} will cause the cluster to lose mass, and two-body relaxation \citep{White1977, Bonnell1998} will drive the cluster to increase its central density and expand its outer layer via negative dynamical heat capacity \citep{LyndenBell1968, LyndenBell1999}. External effects in the form of tides can strip the stars from the outer layers of the cluster if the tidal field is sufficently strong. \citet{Giersz1994} and \citet{Baumgardt2002} showed that for clusters evolving in isolation, the dissolution time-scale could be up to 1000 times the initial relaxation time. It is therefore the galactic tidal environment which has the greatest effect on the lifetime of a cluster \citep{Gieles2008}. 
       
The galactic environment affects clusters through tidal forces that can be sorted by their time-scales. If the change in tidal forces affecting the cluster occurs over a period which is longer than the internal evolution of the cluster, the tides are referred to as adiabatic \citep{Renaud2018}. The majority of globular clusters reside in the halo, far from their galactic centre \citep{Gratton2019}, and the tidal effects on these can be described by adiabatic tides which can be calculated by assuming the galaxy as a point mass (see  \citealp{Chernoff1990} or \citealp{Wang2016}). Further, if a galactic bar is present, clusters in the vicinity will be tidally affected. \citet{Berentzen2012} have shown that a bar will induce periodic expansion and contraction of the outer parts of the cluster. Here, the dissolution time-scale for the cluster is primarily set by the average tidal force experienced along the orbit. 

Interactions and mergers between galaxies can also affect the tidal field \citep{Renaud2018}. In many cases, these interactions can trigger a significant increase in compressive tides over large volumes which can increase the star and cluster formation of a galaxy, but also permenantly alter the mass distribution and therefore the galactic tidal field \citep{Renaud2008, Renaud2009}. These compressive tides are relatively short-lived and last between 30 to 100 Myr \citep{Renaud2009}.  

A cluster can also experience tidal perturbations which occur over much shorter time-scales in the form of tidal shocks. These rapid changes in the galactic tidal field can be produced by spiral arms, the disc, or an encounter with a giant molecular cloud (GMC). \citet{Gieles2007} found that if the passage through a spiral arm is shorter than the stellar crossing time, a spiral arm can produce tidal shocks. However, these shocks have little effect on the overall mass loss of the cluster, since most of the energy gained goes into a few high velocity stellar escapers in the outer part of the cluster. Passages through the galactic disc will produce compressive tides on the cluster in the form of disc shocks \citep{Ostriker1972}. The energy gained by a cluster is greater for disc shocks which occur closer to the centre of the galaxy or through the spiral arms where the surface density is much higher. Clusters which go through the disc with a low speed (i.e. orbits near the disc) or with a high orbital inclination to the disc will result in weaker tidal shocks, whereas clusters at several 100 pc above the galactic plane will experience the most extreme disc shocks \citep{Martinez2017}. 

\citet{Spitzer1958} analytically found that tidal shocks from GMCs can tidal heat a cluster to a point where the cluster is destroyed on a time-scale which scales with the cluster density. Simulations by \citet{Gieles2006} found that the dissolution time of clusters impacted by GMCs in the solar neighbourhood is a factor of 3.5 shorter than the dissolution from adiabatic tides. GMCs are therefore an important factor when it comes to understanding the disruption and destruction of stellar clusters.    
   
In this paper, we investigate the tidal effects of GMCs on a population of young clusters in M51, by combining a galactic model of M51 with $N$-body simulations of 5000 individual clusters over a period of 200 Myr. M51 is a perfect labatory to test the evolution of stellar clusters because of its relative proximity of $8.58\pm0.10$ Mpc \citep{McQuinn2016} and its nearly face-on orientation \citep{Tully1974,Shetty2007,deBlok2008, Colombo2014b}. Here, not only can the gas distribution be tracked \citep{Schinnerer2013}, but also individual GMCs can be resolved \citep{Colombo2014} which provides information about their physical properties. Several investigations of the stellar cluster population have been made \citep{Gieles2009, Chandar2016, Messa2018, Messa2018b} which provide us with a reference frame to compare our results with. 

Our study is motivated by the papers of \citet{Messa2018} and \citet{Messa2018b} (hereafter  \citetalias{Messa2018} and \citetalias{Messa2018b}) which have made detailed studies of the mass and age function of the cluster population of M51 and how these functions change across different regions of the galaxy. The investigation of \citetalias{Messa2018} and \citetalias{Messa2018b} mainly focuses on the higher mass spectrum of the cluster mass function, whereas we are more focused on the lower mass end, since these clusters will be more sensitive to disruption by GMCs.           
 
The paper is constructed in the following way: Section \ref{sec:M51model} describes the galactic model of M51, the implementation and distribution of GMCs, and our best estimate of the initial velocity dispersion. In Section \ref{sec:Simulations}, we explain the initial setup of clusters in the galactic model and $N$-body simulations. We discuss and compare our cluster population to observations in Section \ref{sec:Results}. In Section \ref{sec:Results2}, we investigate the impact of GMC encounters on the clusters and end with our conclusions in Section \ref{sec:Conclusion}.

%The evolution of a stellar cluster is therefore not only a result of internal and but also external effects. The internal effects mainly constitues stellar evolution and two-body relaxation. Stellar evolution will drive mass from the cluster and effectively weaken the gravitational potential of the cluster which is most important for younger clusters where stellar feedback is most extreme. Two-body relaxation will drive the stars in the clusters towards equipartition which leads to mass segregration \citep{White1977, Bonnell1998}. This causes energy to be transfered from the inner parts to the outskirts of the cluster which leads to an expansion of the outer parts of the clusters. The cluster will expand until a point in which the external effects in the form of tidal forces will equal that of the cluster. 
%transports energy from the center of the cluster to its outer layers  

%M51 is one of the most well known grand spiral galaxies. With it having a nearly face on view it it is a perfect labotory for studing the distribution of GMCs and stellar clusters.  
%- M51 observation history\\
%- clusters and GMCs in M51\\
%- Goal of the paper and why is it interesting to look at\\
%- Short description of method\\
%- Paper organised as follows:

\section{M51 Model}
\label{sec:M51model}
The model of M51 consists of an axisymmetric potential, two spiral arms whose pattern evolves over the duration of the simulation, and GMCs. During the evolution of the galactic model, the GMCs are being born and destroyed continuously. 
                                                      
The axissymmetric potential consists of a bulge, a stellar disc, a gas disc, and a dark matter halo. Both the bulge and dark matter halo are represented by a spheroidal \citet{Hernquist1990} profile, given by
\begin{equation}
\rho_{\mathrm{spheroid}}(r) = \frac{M_{\mathrm{spheroid}}}{2 \pi} \frac{a}{r(r+a)^3}, 
\end{equation}  
where $r$ is the spherical radius, $a$ is the scale length and $M_{\mathrm{spheroid}}$ is the mass. For the description of the stellar and gas discs, we used a double exponential density profile
\begin{equation}
\rho_{\mathrm{disc}}(R,z) = \frac{\Sigma_0}{2 h_z} \mt{exp} \left ( \frac{-|z|}{h_z} \right ) \mt{exp} \left ( -\frac{R}{h_R} \right ) , 
\end{equation}  
where $h_R$ is the radial scale length, $h_z$ is the scale height, $\Sigma_{0}$ is the disc's central surface density, and $R$ and $z$ are the cylindrical radius and height, respectively. 

Observationally, the surface density of star formation is well correlated with the surface density of molecular gas \citep{Bigiel2008} and we therefore adopt identical scale length and height for the stellar and gas disc. Based on observations by \citet{Schruba2011}, we used a scale length of 2.21 kpc. We chose a scale height of 0.2 kpc, in accordance with hydrodynamical simulations of an isolated model of M51 by \citet{Dobbs2010}. Our scale height is a minimum estimate, since the interaction between M51 and its companion galaxy NGC 5195 is highly likely to have increased the scale height. The total baryonic mass of M51 is estimated to be $(5.8\pm0.1) \times 10^{10}$ \Mo by \citet{Cooper2012}, which we used as a combined mass for the bulge, the two discs, the spiral arms, and GMCs. We chose the mass of the bulge similar to the models of \citet{Tress2020} and \citet{Dobbs2010}, and dark matter halo mass similar to the model of \citet{Dobbs2010}. The mass of the stellar disc is based on the model of \citet{Tress2020}, where we adopt a gas to stellar disc mass fraction of $\sim10$ per cent. The gas in our model consists of the gas in the disc, the spiral arms, and the GMCs. As such, the mass that makes up the spiral arms and GMCs have been subtracted from the gas disc. All the components of the M51 model are listed in Table \ref{table:M51_model}. The masses and scale lengths of the bulge and dark matter halo have been scaled to best reproduce the observed rotation curve of M51 which is observed to be flat out to a radius of about 8 kpc, with a rotation speed of $\sim210$ \kms \citep{Sofue1999,Oikawa2014}. The rotation curve of our galactic model is further discussed in Section \ref{sec:IVD}.

\begin{table}
 \caption{The parameters of the M51 model components.}
 \label{table:M51_model}
 \begin{tabular*}{0.50\textwidth}{@{\extracolsep{\fill}} ccc}
  \hline
  \hline
  & M51 model & \\
  \hline
 \end{tabular*}
 \begin{tabular*}{0.50\textwidth}{@{\extracolsep{\fill}} lccc}
    & Mass & $h_R$ & $h_z$ \\
    & $[\rm{M}_{\odot}]$ & $[\rm{kpc}]$ & $[\rm{kpc}]$ \\	
  \hline
  \\ 
  Dark Matter Halo & $10^{11}$ & 13.45  & - \\
  Bulge & $7\times10^{9}$ & 0.16  & - \\
  Stellar Disc & $4.53\times10^{10}$ & 2.21  & 0.2 \\
  Gas Disc & $8.77\times10^{8}$ & 2.21  & 0.2 \\
  Spiral Arms & $2.55\times10^{9}$ & 2.21  & - \\
  GMCs & $2.24\times10^{9}$ & 2.21  & - \\	
  \hline
  \end{tabular*}
\end{table}

\subsection{Spiral arms}
The grand spiral design of M51 is believed to be caused by the tidal interaction with its companion galaxy (NGC 5195), as shown by \citet{Dobbs2010}. In our model, the two spiral arms are each represented by 100 inhomogenous oblate spheroids with semi-major and minor
axes of 1000 and 500 pc, respectively. The spheroid density decreases linearly as
\begin{equation}
\rho(a) = \rho_0 (1-a),
\end{equation}
where $\rho_0$ is the central density of the spheroid, $a = (x^2/a_0^2 + y^2/a_0^2 + z^2/c_0^2)^{1/2}$, $a_0 = 1000$ pc and $c_0 = 500$ pc. The central density for each spheroid follows the same exponential decrease as the gas disc given by 
\begin{equation}
\rho_0 = \rho_{02} \, e^{-(R-R_s)/h_R},
\end{equation}
where $R_s$ is radius where the spiral pattern starts. $\rho_{02}$ is given by \citet{Pichardo2003} as  
\begin{equation}
\rho_{02} = \frac{ 3 M_s}{2 \pi a_0^2 c_0 \sum_{j=1}^{N} e^{-(R_j - R_s)/h_R}},
\label{eq:rho02}
\end{equation} 
where $M_s$ is the total mass of the spiral arms, $R_j$ is the galactocentric distance of each of the spheroid's centres which is summed over one spiral arm. The total spiral arm mass is $2.55 \times 10^9$ \Mo which is 5 per cent of the total disc mass. To see how the force of each spheroid was calculated, we refer to \citet{Jorgensen2020}. 

The radial span of each spiral arm starts at 1.3 kpc and ends at 10 kpc, with each spheroid having the same radial separation. The spiral pattern is based on hydrodynamical simulations of M51 done by \citet{Dobbs2010}, where a spiral pattern is formed and ends up representing a similar structure to the observed spiral pattern of M51. In order to reproduce the same spiral pattern, we fitted the average pattern speed of the  two arms, $\Omega(R)$, as a function of galactocentric radius to simulation data in Figure 15 of \citet{Dobbs2010}. A best fit with an exponential decreasing function was found to be 
\begin{equation}
\Omega(R) = 92.2 \, e^{-R/1.75} + 22.8 \, \, \, \, \, \, \, \, \rm{km}\,\rm{s}^{-1} \, \rm{kpc}^{-1}.
\label{eq:pattern_speed}
\end{equation}
The data from \citet{Dobbs2010} only span the radial range of $2.5<R<7.5$ kpc, however, $\Omega(R)$ still agrees reasonably with the observation data of \citet{Meidt2008}. For the inner parts of M51, $R<2.3$, \citet{Meidt2008} measured a pattern speed of $96^{-26}_{+16}$ $\rm{km}\,\rm{s}^{-1} \, \rm{kpc}^{-1}$. In this range, $1.3\leq R<2.3$, $\Omega(R)$ goes from 67 to 48 $\rm{km}\,\rm{s}^{-1} \, \rm{kpc}^{-1}$. In the range of $2.3\leq R<3.9$ kpc, $51^{+7}_{-11}$ $\rm{km}\,\rm{s}^{-1} \, \rm{kpc}^{-1}$ was observed. Here, $\Omega(R)$ ranges from 47 to 33 $\rm{km}\,\rm{s}^{-1} \, \rm{kpc}^{-1}$. In the range of $3.9\leq R \leq 5.3$ kpc, observations are $23^{-7}_{+6}$ $\rm{km}\,\rm{s}^{-1} \, \rm{kpc}^{-1}$, where $\Omega(R)$ ranges from 32 to 27 $\rm{km}\,\rm{s}^{-1} \, \rm{kpc}^{-1}$.          

Based on Eq. \ref{eq:pattern_speed}, each spheroid was given its own pattern speed which means that the spiral structure will wind up, as time evolves. The simulations of \citet{Dobbs2010} occur over a time period of 300 Myr until the present time, whereas our simulations of M51 start 100 Myr later. To set up our spiral pattern, we assumed that each arm starts out as a straight structure and evolved them for 100 Myr to get the initial positions in our model.

\subsection{Number of GMCs}
In order to evaluate and create a realistic environment that represents M51, we used the GMC catalogue constructed by \citet{Colombo2014} which include 1507 individual GMCs. The identified parameters in the catalogue includes positions (right ascension and declination), virial masses, viral radii, and radial velocities. The observations for this catalogue was obtained by the PdBI Arcsecond Whirlpool Survey (PAWS; \citet{Schinnerer2013}), with the Plateaue de Bure Interferometer (PbBI) and IRAM 30 m telescope\footnote{For a detailed description of the data reduction, see \citet{Pety2013}}. The PAWS covers the inner disc of M51 and has a field-of-view (FoV) of $\sim 270^{''} \times 170^{''}$. This area does not cover the entire surface of M51 and thus the total number of GMCs is likely higher than 1507. In order to make a rough estimate of the total number of GMCs in M51, we first projected the observed GMCs down into an $x-y$ plane. The plane is the face-on view of M51 which we constructed by using a position angle, P.A. = $178\pm3^{\circ}$, and inclination, $i = 22\pm5^{\circ}$, estimated by  \citet{Colombo2014b}. We also adopted the same location for the centre of M51 as \citet{Colombo2014b} for consistency. The coordinates of the centre, $(x_0,y_0) = (13^{\rm{h}}29^{\rm{m}}52\overset{\rm{s}}{.}71, 47^{\circ} 11^{'}42\overset{''}{.}79)$, are measured by \citet{Hagiwara2007} by observations of $\rm{H}_2\rm{O}$ maser emission associated with the nucleus of M51. We adopted a distance to M51 of $8.58\pm0.10$ Mpc, measured by \citet{McQuinn2016}. 

The PAWS FoV is complete out to $\sim 3.5$ kpc but continues out to $\sim 8.5$ kpc. We divided this area between 3.5 and 8.5 kpc into annuli with sizes of 0.1 kpc. For each annulus, we calculated the average surface density of GMCs in the area covered by the PAWS FoV. Assuming a constant GMC surface density for each annulus, we used a monte carlo scheme to account for the missing area not covered by the PAWs FoV. We estimate that the total number of GMCs in M51 is around 2300. This is probably at the higher end of the actual number of GMCs, since the PAWS FoV includes the spiral arms, however, our M51 model does also extend out to 10 kpc and not the 8.5 kpc which was covered by the PAWS. 
 
The GMC mass function has been estimated by \citet{Colombo2014} and follows a truncated power-law \citep{Williams1997}. The cumulative mass function is given by
\begin{equation}
N(M' > M) = N_0 \left [  \left (  \frac{M}{M_{\rm{max}}} \right )^{\beta + 1} -  1 \right ],
\label{eq:GMC_PL}
\end{equation}
where $\beta$ is the power-law slope of the differential mass function, $M_{\rm{max}}$ is the maximum mass of the distribution, and $N_0$ is the number of objects more massive than $2^{1/(\beta+1)} M_{\rm{max}}$, which is the mass where the distribution deviates from a power-law. For all observed GMCs, \citet{Colombo2014} found the best estimates for $\beta$, $M_{\rm{max}}$ and $N_0$ to be $-2.29$ , $1.85 \times 10^7$ \Mo and 17, respectively. In our model, we set a minimum GMC mass of $5.5 \times 10^5$ \Mo which accounts for the majority of observed GMCs in M51. As can be seen in Table \ref{table:M51_model}, the amount of mass in the smooth gas disc is quite low. This is because most of the gas in our model has been incorporated into the spiral arms and GMCs.  

\subsection{Evolution of GMCs}
The lifetime of GMCs in M51 was found by \citet{Meidt2015} to be between 20 - 30 Myr. Based on this, we gave each GMC a lifetime of 30 Myr with a parabolic mass evolution over its lifetime of the form
\begin{equation}
M(t) =  4 \left [ - \left ( \frac{t-t_0}{30 \, \mt{Myr}}  \right )^2 +  \frac{t-t_0}{30 \, \mt{Myr}} \right ] \cdot M_0,
\end{equation}
where $M_0$ is the maximum mass of the GMC, $t$ is the current time and $t_0$ is the birth time of the GMC. Each GMC is represented by a Plummer sphere \citep{Plummer1911} with a core radius, $R_c$, which is related to the current mass of the GMC. To find the mass-size relation of the GMCs, we first assumed a $M^{1/2}$ scaling, as described by \citet{Hopkins2012}, and fitted to a subsample of the GMCs from \citet{Colombo2014}. We used the same threshold set by \citet{Colombo2014}, by only using GMCs with a signal-to-noise ratio above 6.5. Furthermore, we also required that the GMCs had estimated masses and radii with relative uncertainties less than 1.0 and 0.5, respectively. Our subsample consist of 183 GMCs with a best core radius fit given by
\begin{equation}
R_c(t) =  10.5 \left( \frac{M(t)}{4.2 \times 10^5 \, \mathrm{M}_{\odot}} \right )^{1/2} \, \mt{pc}.
\end{equation}  

\subsection{Initial velocity dispersion in M51}
\label{sec:IVD}
\begin{figure}
 \includegraphics[width=\columnwidth]{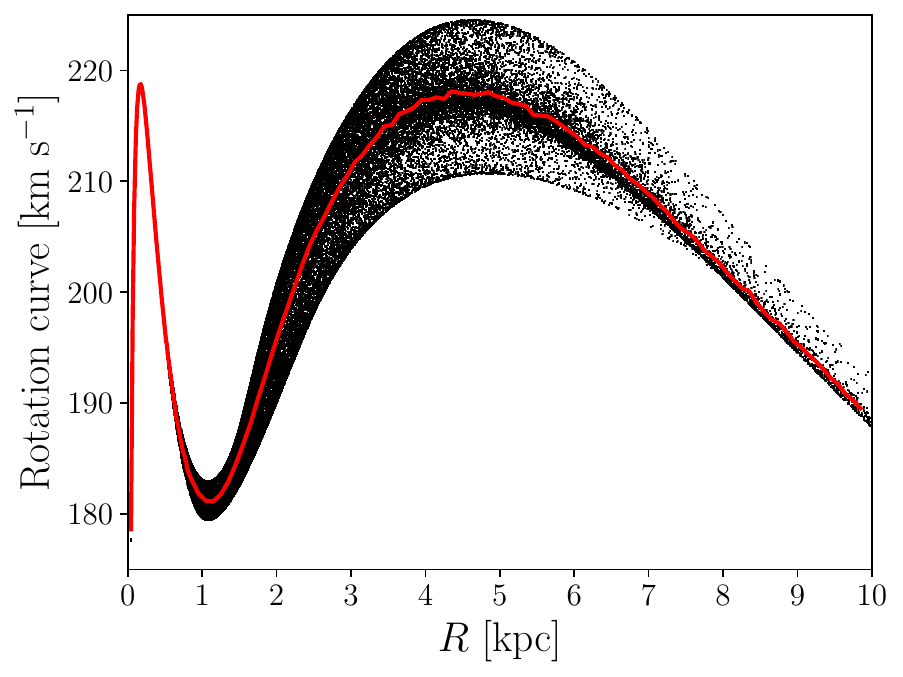}
 \caption{The circular velocity as a function of galactocentric radius. The local circular velocity is affected by nearby spiral arms which cause a spread in the rotation curve. We therefore calculated the local circular velocity for 50 000 test particle (black points) to estimate this spread. The test particles have been binned radially, and an average circular velocity for each bin has been calculated, together with an estimated uncertainty based on the standard deviation. The red line represents the present day rotation curve of our M51 model. The binned data points and associated uncertainties have been omitted for clarity.}
 \label{fig:RC}
\end{figure}

Estimating a correct initial velocity dispersion of GMCs and stellar clusters in M51 is crucial in order to get a realistic GMC encounter history of the stellar clusters. A too low initial velocity dispersion will increase the chance of close GMC encounters, whereas a too high velocity dispersion will spread the GMCs and stellar clusters out into a larger volume and thereby reduce the chance of GMC encounters. In order to estimate the initial velocity dispersion, we used the positions and radial velocities of the GMCs from the catalogue by \citet{Colombo2014}. Due to the short lifetime of the GMCs, the intial velocity dispersion should match the current dispersion. We also assume that the velocity dispersion we find for the vertical component will match the remaining two components. The vertical velocity component, $v_z$, is linked to the observed radial velocity: $v_{\mt{r}}$, the rotation velocity: $v_{\mt{rot}}(R)$ and inclination: $i$, by    
\begin{equation}
v_z = \frac{v_{\mt{r}}}{\mt{cos}(i)} - v_{\mt{rot}}(R) \, \mt{tan}(i) \, \mt{cos} (\theta). 
\label{eq:v_z}
\end{equation}
Here, $\theta$ is the azimuth angle in the disc of a measured point from the major axis, and the rotation velocity is the one from our model which is known for every galactic radius. However, because of the presence of the spiral arms, the local circular velocity will depend on whether the location is close to a spiral arm or not. This will cause a spread in velocity for $v_{\mt{rot}}(R)$, that will fluctuate with $R$. In order to solve this problem, we numerically calculated the average rotation speed at each binned galactic radius by using 50 000 test particles in the phase-space of $0 < R \leq 10$ kpc, $0 \leq \theta < 2 \pi$ and $z=0$. For each radial bin, an estimated uncertainty was calculated based on the standard deviation. The spread of the 50 000 test particles can be seen in Figure \ref{fig:RC}, with a line drawn between the different binned data points which represents the average rotation curve. The binned data points and associated uncertainties have been omitted for clarity. For the inner parts of M51, there is no spread and it is only at $R\sim0.8$ kpc that the presence of the spiral arms starts to affect the rotation curve. The spread in velocities continues to increase from this point until $R\sim4.6$ kpc, where it then decreases out to 10 kpc.   
To calculate $v_{\mt{rot}}(R)$ for each GMC, we applied linear interpolation to the binned data in Figure \ref{fig:RC} which we then used to calculate $v_z$ according to Eq. \ref{eq:v_z}. 

\begin{figure}
 \includegraphics[width=\columnwidth]{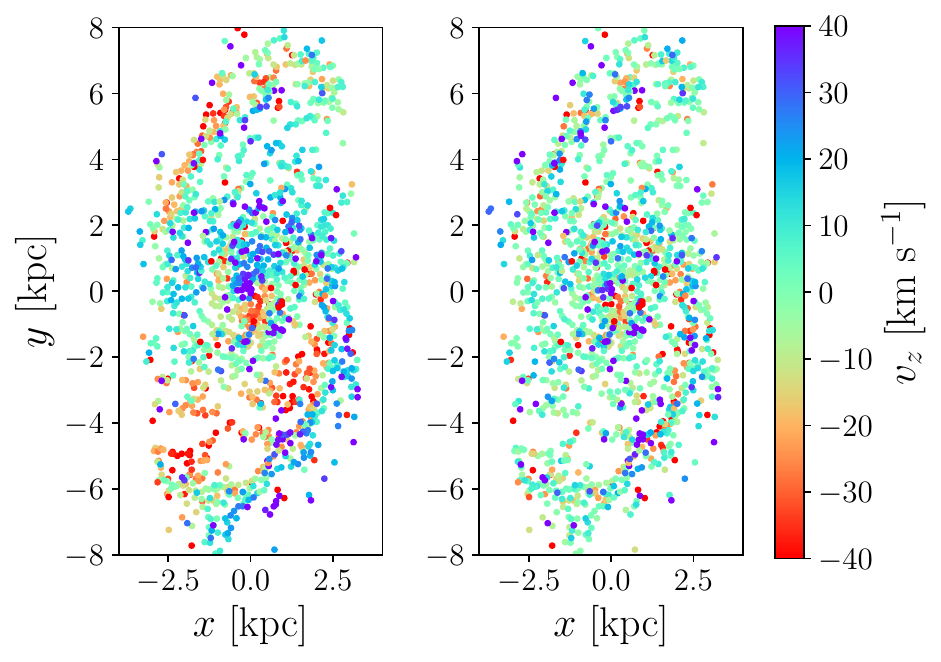}
 \caption{The vertical velocity component of the GMCs in M51. \textit{Left:} Here Eq. \ref{eq:v_z} has been applied to the observed 1507 GMCs from the catalogue of \citet{Colombo2014}. There are several anomalies in the velocity distribution which are likely caused by interaction with NGC 5195. \textit{Right:} Here, an algorithm has been used to remove the anomalies in order to make a more realistic estimate of the initial velocity dispersion of the GMCs in our model. }  
 \label{fig:GMC_Vz}
\end{figure}

The resulting $v_z$ of the GMCs can be seen to the left in Figure \ref{fig:GMC_Vz} in the face-on $x-y$ plane. Each of the 1507 GMCs are colour coded according to their velocity in the range [-40,40] \kms. The GMCs have a velocity distribution of $\sim$ [-150,150] \kms, but we chose a smaller colour code range for clarity. One would assume that the GMCs are born in the disc, $z=0$, with a gaussian velocity distribution which should be mostly independent of where in the $x-y$ plane the GMCs are born. Instead, we see anomalies in the distribution of velocities. There are two groups of anomalies close to the centre of M51, where one group is moving in a negative $v_z$ direction, whereas the other is going in a postive $v_z$ direction. We also see traces of this in the two spiral arms. For the lower spiral arm, we see a majority of postive velocities, together with an area between the centre and the arm, where the majority of GMCs have negative velocities. The upper spiral arm is more dominated by GMCs with negative velocities and no other significant structures. These anomalies are most likely caused by the interaction with NGC 5195, which is also responsible for the observed warp and bend in the disc of M51. \citet{Oikawa2014} reported that the disc is nearly flat in the inner part with a constant inclination angle, however, at a galactic radius of 7.5 kpc it bends by about 27 degrees. 

In order to remove the majority of these anomalies, which otherwise would give us an extremely high initial velocity dispersion for the GMCs in our model of M51, we divided the GMCs into a $20\times9$ grid in the $x-y$ plane. For cells containing 5 or more GMCs within them, the median $v_z$ was calculated. For cells with less than 5 GMCs, we used a nearby neighbour search algorithm to calculate the median velocity based on the neighbouring cells. Afterwards, the velocity cell was subtracted from the original velocities which can be seen to the right in Figure \ref{fig:GMC_Vz}. Most of the anomalies have been removed, even though there are still some remnants of them in the centre of M51.

To find the initial velocity distribution for our model of M51, that would best reproduce the calibrated observations, we ran several simulations of our model. Each simulation had a different initial gaussian velocity distribution, $\sigma_i$, for the GMCs. The spatial distribution of the GMCs is addressed in Section \ref{sec:BirthGMC}, however, changing the birth locations had no significant effect on the resulting vertical velocity distribution. We found that having an $\sigma_i$ of 10 \kms which is similar to the Milky Way \citep{Holmberg2009} could not reproduce the observed spread in M51. Instead, we found that a $\sigma_i$ which best fit the observation to be $\sim 35$ \kms. In our model, both stellar clusters and GMCs are born with an initial gaussian velocity dispersion of $\sigma_{i}$ in each of the three spatial directions. 
\begin{figure*}
 \includegraphics[width=1.0\textwidth]{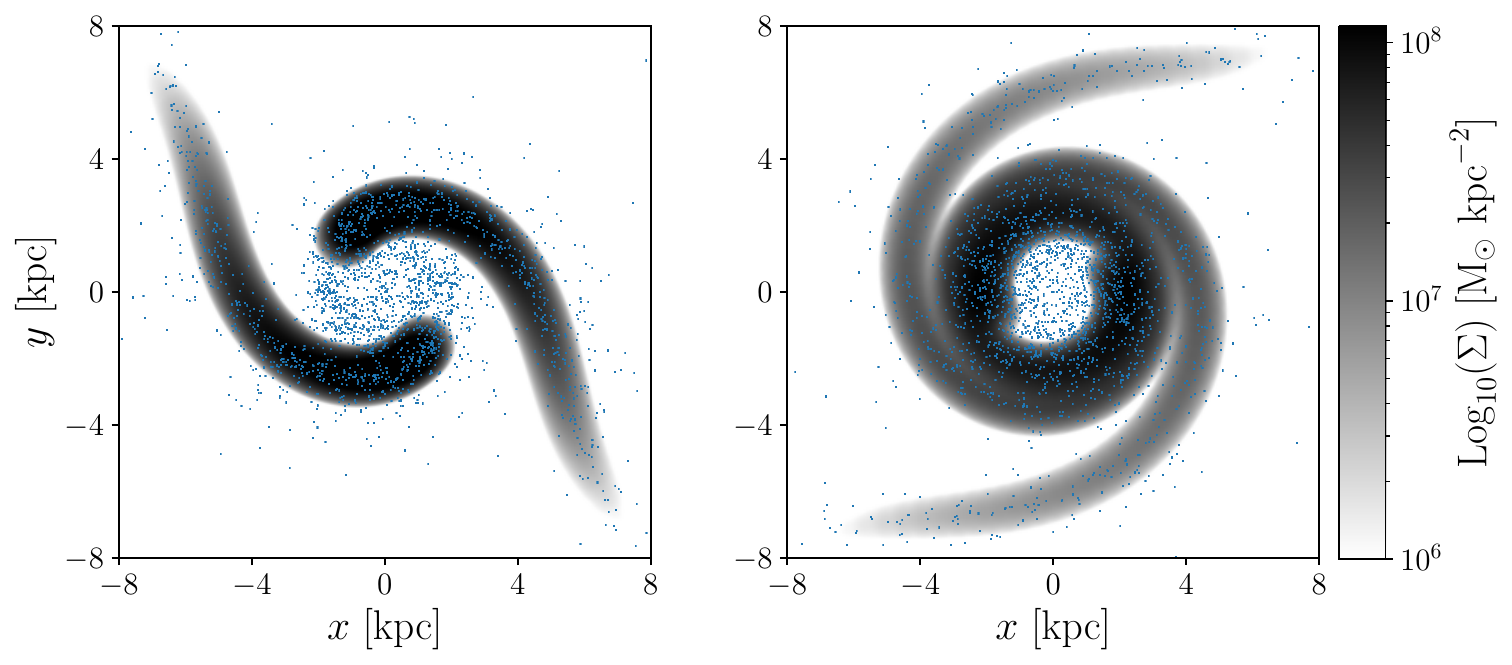}
 \caption{The evolution of the spiral pattern in the galactic plane from the start of the galactic simulation (\textit{left}) to the end of the simulation after 200 Myr (\textit{right}). The spiral arms are represented by their surface density, $\Sigma$, and the current distribution of the GMCs is represented by blue dots. The GMCs born at a galactocentric radius within the spiral structure at 1.3 kpc are uniformly distributed. The GMCs born beyond this radius follow the radial density distribution of the spiral arms. 90 per cent of these GMCs are born in the spiral arms and the rest are born at randomly distributed locations in the galaxy.}
 \label{fig:pattern}
\end{figure*}

\subsection{Birth location of GMCs}
\label{sec:BirthGMC}
Because the lifetimes of GMCs are very short, the GMCs are born very close to their current observed positions. Following \citet{Colombo2014}, we adopted a central region defined to be within the galactocentric radius of 1.3 kpc for which 300 out of our 2300 GMCs are born with a uniform distribution. The remaining 2000 GMCs follow the same radial distribution as the density of the spiral arms with a scale length of 2.21 kpc. GMCs are normally associated with the spiral arms of galaxies where we expect them to be born. \citet{Colombo2014} found that a significant fraction of GMCs ($\sim$ 44 per cent)  is located in inter-arm (IA) region of M51. However, the IA region only contains a few GMCs with masses greater than $10^{6.5}$ \Mo and as such, the most massive GMCs are found in the spiral arm (SA) region. \citet{Koda2009} have argued that the majority of GMCs found in the IA region of M51 could not have formed in situ, but they would instead have been born in the spiral arms and then later crossed into the IA region. \citet{Koda2009} based their arguement on the time-scale it would take to accumulate enough mass in the IA region to form a GMC which they found to be of the order $\sim 100$ Myr. Given the large velocity dispersion observed in M51, it is indeed possible for a GMC with a lifetime of $\sim 30$ Myr to be born in the spiral arms and move into the IA region before it is dispersed. However, the calculation done by \citet{Koda2009} assumed a velocity dispersion of 10 \kmso for the gas in the disc. The time-scale for mass growth is $\propto \sigma^{-1}$ which means that adopting a velocity dispersion of 35 \kms, reduces the time-scale down to $\sim 28.6$ Myr which indicates, that it could be possible for a fraction of GMCs to form in situ in the IA region. Because of this, we allowed a small fraction, i.e. 10 per cent, of the 2000 GMCs to be born at random locations independent of the positions of the spiral arms. All of the remaining GMCs are born within 50 pc of the locus of the spiral arms. 

The SA and IA regions can be defined in various ways. \citet{Colombo2014} defined the regions by the kinematic nature of the CO gas, whereas \citetalias{Messa2018b} defined the SA region based on V-band brightness. In our case, we make the simple assumption that the SA region is within 250 pc of the spiral locus projected onto the galactic plane. By doing this, we found that at the present time $\sim 37$ per cent of GMCs are located in IA region in our model which corresponds well with the results of \citet{Colombo2014}, which observed a value of $\sim 42$ per cent for GMCs with masses larger than $5.5 \times 10^5$ \MO. The location of the GMCs can be seen in Figure \ref{fig:pattern}, together with the initial and final spiral pattern of our M51 model. The spiral arms are represented by their surface density, $\Sigma$.

\subsection{The companion of M51}
\label{sec:M51b}
M51 is currently interacting with its companion galaxy, NGC 5195. By using hydrodynamical simulations, \citet{Dobbs2010} and \citet{Tress2020} have shown that this interaction is the cause for the formation of the grand spiral arms in M51. The two galaxies are currently merging, and thus it is necessary to consider the tidal effects that NGC 5195 might have on the stellar clusters in M51. In their simulations, \citet{Tress2020} used a mass of $4 \times 10^{10}$ \Mo to represent NGC 5195. At the pericentric passage in their orbit, the two galaxies were seperated by a distance of 14 kpc. Assuming stellar clusters out to a galactic distance of 6 kpc, the minimum distance between the clusters and NGC 5195 would be 8 kpc. Considering a typical GMC encounter with a GMC mass of $\sim 10^6$ \Mo in our model, with a separation of 100 pc, which is a fairly wide encounter, the tidal field strength of NGC 5195 is only 7.8 per cent of the GMC encounter. We therefore found it reasonable to ignore the tidal effects of NGC 5195 in our model.   

\section{Simulations}
\label{sec:Simulations}
The evolution of a stellar cluster, especially a low-mass cluster, is dependent on its local environment and it is therefore important to know what kind of external perturbations the cluster experiences throughout its lifetime. Clusters that spend much of their time closer to the galactic centre will be subject to stronger tidal forces, which will reduce their lifespan. Tidal forces from the presence of spiral arms and encounters with GMCs will further create unique tidal evolutionary histories for each cluster, which in the end will dictate the resulting $N$-body simulation. We are specifically interested in how big an impact the GMC encounters have on the clusters. To investigate this, we run two $N$-body simulations for each unique cluster: one where the tidal effects of the GMCs are present, and one where this effect is ignored. We make a direct comparison for each cluster and the differences will be a result of the tidal effects from the GMCs. In the following sections, we will refer to the realistic clusters, the ones which will feel the tidal effect of GMCs, as C$_{\rm{R}}$ clusters. The clusters which are not affected by the tidal effects of the GMCs will be referred to as C$_{\rm{N}}$ clusters. 
\subsection{Galactic simulation and cluster setup in M51} 
%\subsection{Stellar cluster orbits and GMC encounters in M51 model}
The simulation of M51 starts out 200 Myr in the past and continues up until the present time. We ran the simulation with 10000 clusters which were represented as test particles, all of which followed the same spatial distribution as the GMCs. The simulation was done using a 4th order Runge-Kutta integrator with a time step of 0.1 Myr. The interaction between M51 and its companion started around 300 - 500 Myr ago \citep{Salo2000b,Dobbs2010, Tress2020}, and we therefore assume that the cluster formation rate has remained constant during the last 200 Myr. In order to directly compare our results to the cluster observations of \citetalias{Messa2018, Messa2018b}, which had a lower cluster detection limit at 1.3 kpc, we required that the galactocentric radius of the clusters in our sample to be greater than 1.3 kpc at the end of the simulation. We also set an upper galactocentric radius limit of 6 kpc due to the potential tidal effects of the companion of M51, as discussed in Section \ref{sec:M51b}. Out of the 10000 clusters, we randomly selected 5000 which fulfilled our criterion of $1.3\le R \le 6$ kpc.

\subsection{$N$-body Simulation}
For the  $N$-body simulations of the stellar clusters, we used the code \textsc{nbody6tt} by \citet{Renaud2011} which is a modified version of the code \textsc{nbody6} \citep{Aarseth2003}. \textsc{nbody6tt} allows the use of a tidal tensor which can be used to describe the local galactic environment of a stellar cluster. The tidal tensor is defined as
\begin{equation}
\boldsymbol{T}_{ij} = -\frac{\partial^2 \Phi}{\partial x_i \partial x_j},
\label{eq:tt}
\end{equation}
where $\Phi$ is the local galactic gravitational potential. We constructed two tidal tensors for each cluster in the M51 model: one realistic tidal tensor (C$_{\rm{R}}$) and a tidal tensor where the tidal contribution from the GMCs had been removed (C$_{\rm{N}}$). The tidal tensor was created by numerically differentating the force acting on the cluster for each time step within a cube of size $2\delta$. We chose a $\delta$-value of 15 pc which means that most of the galactic environment can be described by the tidal tensor. The only exception to this happens when the cluster is experiencing close GMC encounters that create strong tides, which we will address in Section \ref{sec:CE}.

\subsection{Initial cluster setup}
\label{sec:Initial_cluster}
The initial cluster mass function in M51 has been observed to have a slope of $-2$ by \citet{Chandar2016} and \citetalias{Messa2018} which means that a larger fraction of cluster will have relatively low masses. For a robust investigation of specific cluster mass ranges, we need to simulate a certain number of clusters. We chose to investigate the mass range $[6000 - 24000]$ \Mo, by simulating 1000 unique clusters. Following an initial cluster mass function of $-2$, in order to investigate the cluster population down to cluster masses of $600$ \MO, we would need to perform an additional 12000 unique cluster $N$-body simulations. Due to computational limitations and time saving, we only investigated 4000 unique clusters in the mass range $[600-6000]$ \MO. Hence, the 5000 clusters were divided into two different mass ranges which both follow the same initial cluster mass function with a slope of $-2$. When the total cluster population is analysed a normalisation is performed.

For the initial cluster half-mass radius we constructed a probability density function based on the observations of the size distribution of clusters in M51 by \citet{Chandar2016}. For clusters with masses in the range of $10^{4.5}$ to $10^{4.8}$ \Mo and ages less than 10 Myr, \citet{Chandar2016} found clusters with effective radii distributed between 1 to 6 pc. From this, we fitted an initial cluster radius function (ICRF) with a best-fitting of 
\begin{equation}
\rm{ICRF}(r) = \xi_0 \, e^{-r/\xi_1} + \xi_2,  \quad  r \in [1,6] \,\, \rm{pc},
\label{eq:ICRF}
\end{equation}   
where $\xi_0 = 2.08$ pc, $\xi_1 = 0.78$ pc and $\xi_2 = 0.11$ pc. \citet{Chandar2016} found no obvious relation between cluster mass and radius for clusters with ages younger than 10 Myr, which we also assumed to be the case for our clusters. It should be noted that our ICRF is consistent with the 429 M51 cluster radii, with cluster ages less than 10 Myr in an identical mass range as our clusters, from the cluster catalogue by \citet{Brown2021}, who measured the effective radii of clusters in 31 different galaxies from the Legacy Extragalactic UV Survey (LEGUS). 

For the $N$-body simulations, the stars were distributed with the \citet{Kroupa2001} initial mass function with stellar masses in the range 0.1 - 50 \Mo and spatially distributed according to a \citet{Plummer1911} distribution. All stars were born as single stars with stellar evolution turned on and a metallicity of 0.02. 

\subsection{Cluster escapers}
\label{sec:CE}
Due to two-body relaxation and equipartition, the high-mass stars will move towards the cluster centre, whereas the low-mass stars will diffuse towards the outer parts of the cluster. As this process continues, the cluster will fill its tidal radius, and stars with positive energies that cross this barrier will be able to escape the cluster. The tidal radius, $r_t$, is defined by the maximum eigenvalue, $\lambda_{\rm{max}}$, of the tidal tensor \citep{Renaud2011} as
\begin{equation}
r_t = \left ( \frac{G M_c}{\lambda_{\rm{max}}} \right )^{1/3},
\label{eq:rt}
\end{equation}
where $M_c$ is the mass of the cluster and $G$ is the gravitational constant. At certain points in a cluster's evolution when it either interacts with a spiral arm, experiences a disc shock, or a GMC encounter, all three eigenvalues of the tidal tensor can become negative. This means that only compressive tides are acting on the cluster and that the tidal radius cannot be defined. As such, instead of using a radial cut for when stellar escapers should be removed in \textsc{nbody6tt}, we instead evaluate the escaped and bound stars at the end of the $N$-body simulation for each cluster. This is done by using Eq. \ref{eq:rt} and evaluating the time closest to the end of the simulation for which there exists a positive eigenvalue in the tidal tensor. We remind the reader that we have two different tidal tensors for each cluster: One containing the contribution of the GMCs (C$_{\rm{R}}$) and one where this effect has been removed (C$_{\rm{N}}$). In order to make the best comparison between the two different cluster versions, we calculated the tidal radius using the tidal tensor without the contribution of the GMCs for both C$_{\rm{R}}$ and C$_{\rm{N}}$ clusters. We defined a cluster to be destroyed if it had less than 200 stars at the end of the $N$-body simulation. At this low number of stars, the cluster will have a mass of $\sim 100$ \Mo and be too faint to be detected in cluster catalogues of M51.

\subsection{Recreating GMC encounters in \textsc{nbody6tt}}
The tidal tensor provides a linear tidal force and thus this approximation begins to break down if an object which produces a strong tidal field comes too close to a stellar cluster. This happens when a cluster has a strong GMC encounter where the distance of closest separation between the cluster and the GMC is below 60 pc. A strong encounter can be defined using the approximation for the fractional change in kinetic energy of the cluster, described by \citet{Gustafsson2016} as 

\begin{equation}
\delta_E = \frac{8 G M^2 r_h^3}{3 M_c b^4 V^2}.
\label{eq:dE}
\end{equation}
Here, $M$ is the mass of the GMC, $r_h$ is the cluster half-mass radius, $b$ is the impact parameter and $V$ is the relative velocity between the two objects at infinity. \citet{Gustafsson2016} showed that encounters which generate $\delta_E \lesssim 0.01$ have little to no effect on the evolution of a cluster, and we therefore defined all strong GMC encounters to be encounters which have a separation at closest approach below 60 pc and $\delta_E \ge 0.01$. There are still GMC encounters with large $\delta_E$ at further distances, but these are well approximated in the tidal tensor. Similarly, the GMC encounters with $\delta_E < 0.01$ are still included in the tidal tensor, but have little effect on the cluster compared to the rest of the galactic environment. 

During a strong GMC encounter, we removed any GMC contribution to the tidal tensor and instead directly simulated the GMC encounter in \textsc{nbody6tt}. To do this, we first had to recreate the encounter between the cluster and GMC to calibrate the initial impact parameter between the two objects. This is necessary because our M51 model and \textsc{nbody6tt} do not share the same reference frame and the GMCs do not experience the galactic potential in \textsc{nbody6tt}, since the galactic potential is described by the tidal tensor which only affects the cluster. Furthermore, the GMCs do not feel the forces of other GMCs in \textsc{nbody6tt} which they do in our M51 model. The recreation of the orbit for each strong encounter was done using a 4th order Runge-Kutta integrator, which starts the integration of the GMC orbit at the point of closest approach and then integrates the orbitial path of the GMC to find the position and velocity of the GMC at birth. The recreation of an orbit is straight forward if the cluster only experiences one GMC encounter over the lifetime of the specific GMC. However, for multiple GMC encounters during the same time period, the orbital integration is affected by the chaotic nature of the many-body problem. It can therefore be very difficult to recreate all the orbits of the GMCs with perfect precision. The computed impact parameter, birth position, and velocity relative to the cluster of each GMC to be used in \textsc{nbody6tt}, was accepted if the distance between the cluster and GMC at closest approach had an error lower than 10 per cent. The time step for the integration was initially set to 0.01 Myr, but would be decreased to 0.0001 Myr in some cases to get acceptable initial conditions for the GMCs of certain clusters. This would normally happen if a cluster experienced multiple GMC encounters at the same time.

\begin{figure}
 \includegraphics[width=\columnwidth]{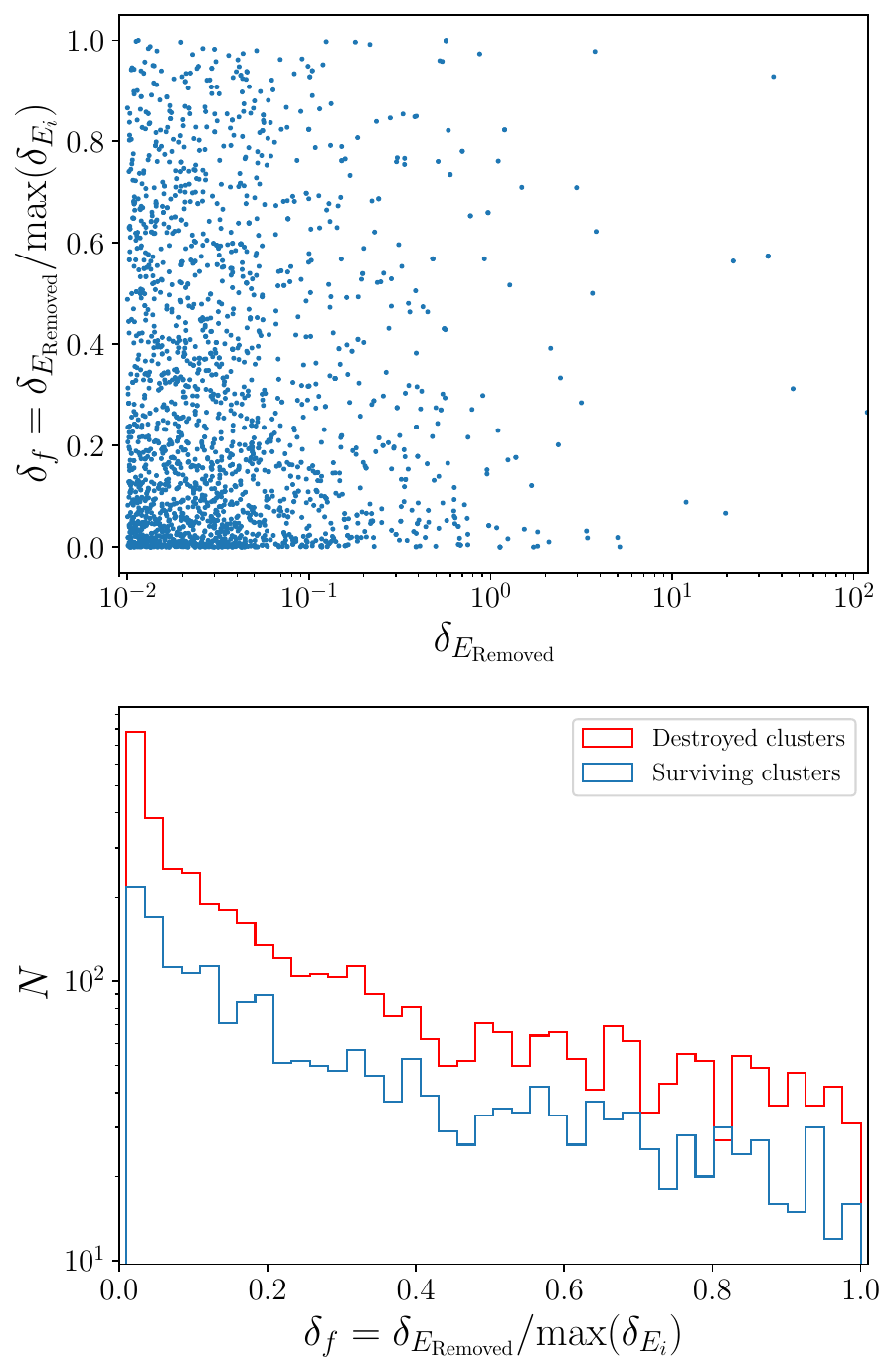}
 \caption{\textit{Top:} The energy ratio, $\delta_f$, of the strongest GMC encounter $\rm{max}$$(\delta_{E_i})$ and the removed GMC encounter $\delta_{E_{\rm{Removed}}}$ for clusters which have serveral GMC encounters in the same period of time, as a function of $\delta_{E_{\rm{Removed}}}$. Here, only the surviving clusters are shown. \textit{Bottom:} The distribution of $\delta_f$ for all 1626 clusters which had to have one or more GMC encounters removed in order to recreate the remaining GMC encounters in \textsc{nbody6tt}.}
 \label{fig:GMC_removed}
\end{figure} 

In some instances, when two or more close GMC encounters happened in close succession, we were not able to produce acceptable initial conditions for the GMCs to recreate the correct orbital encounters, because of the chaotic nature of the many-body problem. In such cases, the GMC encounter with the largest $\delta_{E_i}$ was prioritied with an algorithm that would continuously remove the GMCs which produced the lowest value of $\delta_{E_i}$, until all remaining GMC encounters for the cluster were acceptably reproduced\footnote{Here $i$ represents the different GMCs which are interacting with the cluster at the same time. Meaning that $\delta_{E_{\rm{Removed}}}$ = $\rm{min}$$(\delta_{E_i})$ if the acceptable orbital encounters could not be reproduced.}. This does, however, also mean that for some clusters, the impact of some GMC encounter histories are underestimated because we were not able to reliably reproduce the encounters. We had 1626 clusters which required the removal of one or more GMC encounters. We denote the energy of the removed GMC encounter as $\delta_{E_{\rm{Removed}}}$. 783 of the 1626 clusters were destroyed by the end of the simulations and therefore the exclusion of their associated GMC encounters are unlikely to change the final outcome for these clusters. The 843 clusters which did survive can be seen in the top of Figure \ref{fig:GMC_removed}, where $\delta_f$ is shown as a function of $\delta_{E_{\rm{Removed}}}$. Here, $\delta_f$ is the ratio between $\delta_{E_{\rm{Removed}}}$ and the strongest GMC encounter, $\rm{max}$$(\delta_{E_i})$, i.e. $\delta_f = \delta_{E_{\rm{Removed}}}/\rm{max}$$(\delta_{E_i})$. The majority of the removed GMC encounters have a $\delta_f$ below 0.2, meaning that a much more energetic GMC encounter is dominating the dynamics of the cluster at a similar point in time, and that the removal of these GMC encounters has little effect on the perturbation of the clusters. There are some clusters which have a high value of $\delta_f$, but most of these are low energy encounters which should not affect the clusters too much. There are only 11 GMC encounters which are very energetic ($0.8 > \delta_{E_{\rm{Removed}}}$) and are comparable to the strongest GMC encounters occuring at the same time ($\delta_f > 0.6$) which are related to 9 unique clusters. 5 of these end up with masses that are below the initial cluster mass range and will therefore not be considered in the analysis of the cluster population. The remaining 4 clusters have masses of 1048, 1571, 4116 and 6114 \MO, and the removal of their GMCs encounters are likely to have a high impact on the evolution of these specific clusters. The bottom of Figure \ref{fig:GMC_removed} shows the distribution of $\delta_f$ for both the surviving and destroyed clusters. For all clusters, the majority of removed GMC encounters have less than half the energy compared to another encounter which is affecting the cluster during the same time period. There are many more removed GMC encounters for the destroyed clusters which is simply because these cluster generally have undergone more GMC encounters overall. 

Our method of GMC removal when recreating realistic GMC encounters in \textsc{nbody6tt} will only affect the 843 clusters which survive. Out of these, only 4 clusters have GMC encounters removed which would strongly affect their evolution. Out of all 5000 clusters, this small fraction of clusters should not alter our analysis of the cluster population.

%\section{Results}
\section{The cluster population of M51}
\label{sec:Results}
We have simulated the evolution of 5000 individual clusters where $N$-body simulations have been carried out twice for each cluster: once for a realistic environment with GMCs which we refer to as $C_R$ clusters, and once without the GMCs which we refer to as C$_{\rm{N}}$ clusters. Our analysis is organized in such a way, that we start with a comparision between C$_{\rm{R}}$ and C$_{\rm{N}}$ clusters to the cluster observations made by \citetalias{Messa2018} and \citetalias{Messa2018b}. Then in Section \ref{sec:Results2}, we investigate the differences between C$_{\rm{R}}$ and C$_{\rm{N}}$ clusters to understand the effect that GMCs have on the evolution of clusters in M51. 
\subsection{Mass function}
\begin{figure}
 \includegraphics[width=\columnwidth]{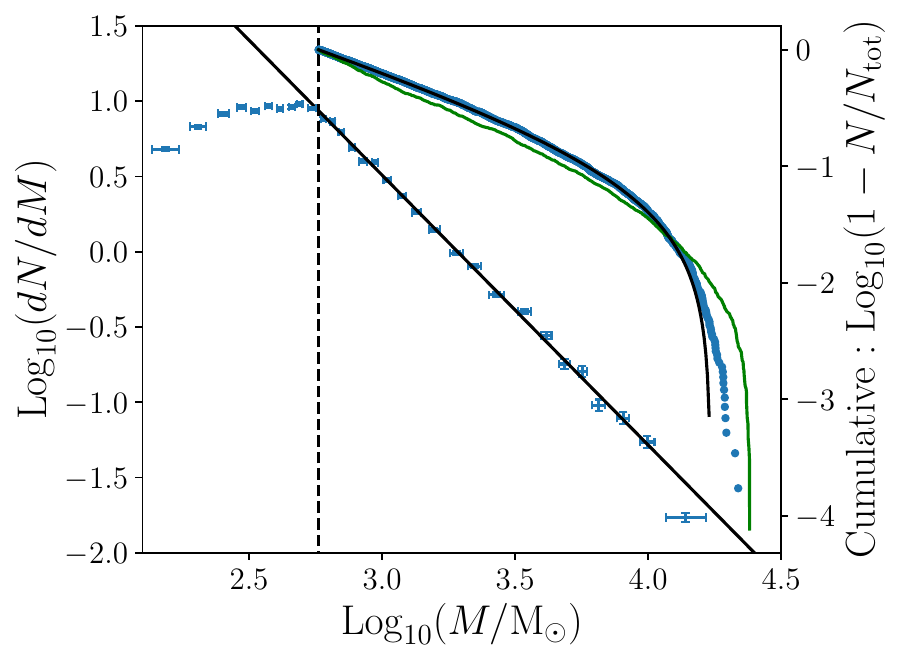}
 \caption{The present day mass function of C$_{\rm{R}}$ clusters. The vertical black dotted line represents the minimum mass cut for the fit, which is the minimum initial cluster mass in our simulations. The cumulative distribution is also shown, where the clusters are represented by the blue dots, the black line is the best fit and the green line represents the initial cluster distribution.}
 \label{fig:MF}
\end{figure}

As mentioned in Section \ref{sec:Initial_cluster}, the intial mass function of our clusters is a power-law with a slope of $-2$. However, the clusters are divided into two groups. Because of this, we want to clarify that the cluster mass function and any other relation regarding our cluster population, which is presented here, will be normalised. The mass function of C$_{\rm{R}}$ clusters can be seen in Figure \ref{fig:MF}, where bins of equal number of clusters have been used. The fit has been made to a single power-law, i.e. $dN/dM \propto M^{\beta}$, for clusters with masses in the range of the initial mass function, where the dashed vertical line indicate the lowest end of the initial mass function. We see a more flattened mass function with $\beta = -1.79\pm 0.02$, compared to the observations of \citetalias{Messa2018}, where they observed no change from an expected initial mass function, with $\beta = -2.01\pm0.04$. 

We find a similar slope for the C$_{\rm{N}}$ clusters which suggests that the overall mass function is not affected by the presence of GMCs over a time period of 200 Myr, but it is rather dominated by the rest of the galactic tidal field. The flattening of the mass function indicates that low-mass clusters are the ones most affected by the galactic tidal field. The majority of our clusters are at a lower mass range than the observations of \citetalias{Messa2018}, which is likely why there is a discrepancy between our results and the observations. If we instead move our mass cut to 5000 \Mo to match that of \citetalias{Messa2018}, we get $\beta = -1.98\pm 0.17$ which is in agreement with the observations to within one standard deviation. The difference between the slopes when applying a mass cut further emphasizes that cluster disruption is mostly affecting the low-mass clusters. All the results of the C$_{\rm{R}}$ and C$_{\rm{N}}$ clusters are listed in Table \ref{table:fits}.    

The binning method can be sensitive to small scale changes, like the dearth of high-mass clusters, as mentioned in \citetalias{Messa2018}. In our simulations, there is a natural dearth of clusters with masses above 24000 \MO. Clusters more massive than this limit would be able to evolve into the mass range we are investigating which could affect the slope of our mass function. To see the small scale changes to the mass function and the truncation of high-mass clusters, we also show the cumulative distribution in Figure \ref{fig:MF}. The cumulative distribution of the C$_{\rm{R}}$ clusters is represented by the blue points, and the green line represents the initial cluster distribution. The black line shows the best fit to a cumulative mass function given by Eq. \ref{eq:GMC_PL}. The low-mass clusters are most affected by the tidal environment which results in an increase in the fraction of clusters compared to the initial distribution in the mass range $10^3$ - $10^4$ \MO. There is a clear truncation at higher masses as suspected, but we find a similar value for $\beta$ if we fit the data to a truncated power-law. We therefore conclude that the binnning method we are using is sufficient to describe the mass function.

The data used in \citetalias{Messa2018} was from the LEGUS where the Wide Field Camera 3 was used, for which the UVIS Channel has a pixel size of $0.04^{''}$. \citetalias{Messa2018} estimated the masses of the clusters using an aperture radius of 4 pixels, together with an aperture correction. The masses of our simulated clusters are based on the energy of each star and the associated tidal radius of the cluster. The observations do not have the same luxury, and we therefore investigated if a fixed aperture applied to our clusters would affect the cluster masses. We used a physical radial cut of 6.7 pc which is comparable to the 4 pixel aperture radius used by \citetalias{Messa2018} before an aperture correction was used. By doing this, we did not find any significant changes in the result of the slope of the mass function. This is because it is largely clusters with masses below our initial cluster mass which are affected, and because the higher mass clusters have most of their mass within this fixed aperture.

\begin{table}
 \caption{Slopes of the mass and age functions which have been fitted to power-laws of $dN/dM \propto M^{\beta}$ and $dN/dt \propto t^{\gamma}$, respectively.}
 \label{table:fits}
 \begin{tabular*}{0.50\textwidth}{@{\extracolsep{\fill}} lcc}
    & C$_{\rm{R}}$ & C$_{\rm{N}}$ \\
  \hline
  %\hline
  Number of surviving clusters   & 3864 & 4058 \\
  \hline
  \hline
  $\frac{dN}{dM}$ \Tstrut\Bstrut & $\beta$ & $\beta$ \\
  %\cmidrule
  \midrule
  %\hline
  Total &               $-1.79\pm 0.02$  & $-1.80\pm 0.03$ \\
  Mass cut: 5000 \MO &  $-1.98\pm 0.17$  & $-2.04\pm 0.12$ \\
  \hline
  \hline
  $\frac{dN}{dM \Delta t}$ \Tstrut\Bstrut & $\beta$ & $\beta$ \\
  %\hline
  \midrule
  1-10 Myr &    $-2.09\pm 0.17$ & $-2.05\pm 0.15$ \\ 
  10-100 Myr &  $-1.85\pm 0.04$ & $-1.89\pm 0.04$ \\
  100-200 Myr & $-1.67\pm 0.04$ & $-1.67\pm 0.05$ \\
  \hline
  \hline
  $\frac{dN}{dt}$ \Tstrut\Bstrut & $\gamma$ & $\gamma$ \\
 % \hline
  \midrule
  Total &  $-0.41\pm0.06$ & $-0.36\pm 0.06$ \\
  $[0.6 - 3.0]  \times 10^3$ \Mo &  $-0.43\pm 0.08$ & $-0.38\pm 0.07$ \\
  $[3.0 - 6.0]  \times 10^3$ \Mo &  $-0.38\pm 0.03$ & $-0.35\pm 0.03$ \\
  $[0.6 - 2.4]  \times 10^4$ \Mo &  $-0.26\pm 0.08$ & $-0.21\pm 0.07$ \\
  $[0.5 - 1.0]  \times 10^4$ \Mo &  $-0.24\pm 0.06$ & $-0.21\pm 0.06$ \\
  $[1.0 - 2.4]  \times 10^4$ \Mo &  $-0.29\pm 0.08$ & $-0.23\pm 0.08$ \\
  \hline
 \end{tabular*}
\end{table}

\begin{figure}
 \includegraphics[width=\columnwidth]{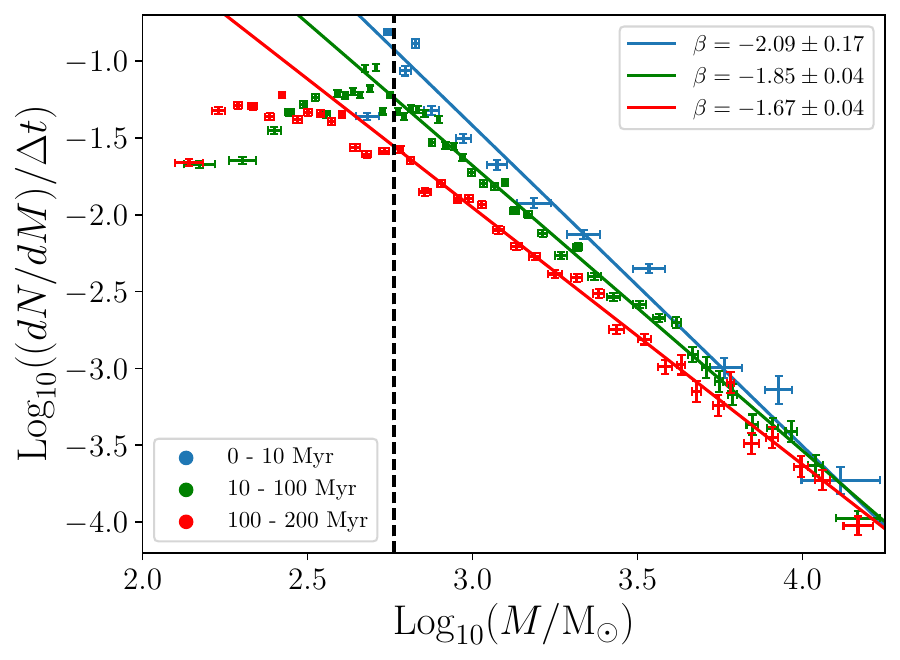}
 \caption{The $C_R$ cluster mass function divided into three different age ranges: 0-10 Myr (blue), 10-100 Myr (green), and 100-200 Myr (red). The black vertical dotted line represents the minimum mass cut for the fits. As the age of the clusters increase, so does the disruption rate which is most significant for the low-mass clusters.}
 \label{fig:MF_age}
\end{figure}

Cluster mass loss and destruction affects the mass function and will depend on the age spread of the cluster distribution. To see how age affects the mass function, we constructed mass functions divided into different cluster age ranges which were then normalised with the age range, i.e. $(dN/dM)/ \Delta t$. We chose similar age ranges as \citetalias{Messa2018}: 0 - 10 Myr, 10 - 100 Myr and 100 - 200 Myr, which for the C$_{\rm{R}}$ clusters can be seen in Figure \ref{fig:MF_age}. For the youngest clusters (blue), we see a clear agreement with the data of \citetalias{Messa2018} who observed $\beta = -2.02\pm0.11$. A similar agreement is found with the observations of \citet{Gieles2009} and \citet{Chandar2016} with $\beta = -2.08\pm0.08$ and $\beta = -2.06\pm0.05$, respectively. The young clusters have not had enough time to evolve or be affected by the galactic tides of M51, and therefore follow the initial mass function with $\beta = -2.09\pm 0.17$. As the clusters get older, we see a flattening in the slope of $\beta = -1.85\pm 0.04$ and $\beta = -1.67\pm 0.04$ for the age ranges of 10 - 100 Myr and 100 - 200 Myr, respectively. This indicates that the low-mass clusters are much more sensitive to the time spent in the galactic field compared to clusters with higher masses. \citetalias{Messa2018} only find a significant difference from the initial mass function for clusters in the age range of 100 - 200 Myr, with a slope of $\beta = -1.85\pm 0.06$. Our different results are likely to be the result of the mass cut used for the clusters. \citetalias{Messa2018} found a similar issue when comparing their results to earlier results of \citet{Chandar2016} and \citet{Gieles2009}, where a reasonable agreement was found for clusters with ages up to 100 Myr, but not for older ages. The most likely reason for a discrepancy between results at older ages was attributed to the completeness limit, where \citet{Chandar2016} and \citet{Gieles2009} had a mass cut in their older ages of $6\times 10^4$ \Mo and $\sim 10^4$ \MO, respectively. It should also be noted that determining reliable cluster ages can be difficult because of the age and dust reddening degeneracy problem. Young dust-reddened clusters may appear older, or older clusters might be classified as younger sources as described by \citet{Whitmore2023}. This misclassification of the cluster ages can lead to incorrect conclusions about a cluster population.

We do not see a significant difference in $\beta$ in any of the age ranges between C$_{\rm{R}}$ and C$_{\rm{N}}$ clusters, indicating that the effects of GMC encounters on the cluster population cannot be detected in the mass function.

\subsection{Age Function}
\label{sec:age_function}
The inferred slope of the mass function can depend on the cluster mass interval which is used. Because of this, it can be difficult to compare observations for different observational cluster mass ranges. It is therefore more useful to compare the age function of a cluster distribution for which the same mass range is observed. Due to computational restrictions and our interest in the lower-mass population, our cluster mass ranges are lower than the bulk of the observed clusters. However, we are able to make a direct comparison to the observations of \citetalias{Messa2018} for clusters with masses between $5 \times 10^3$ and $10^4$ \MO. 

\begin{figure}
 \includegraphics[width=\columnwidth]{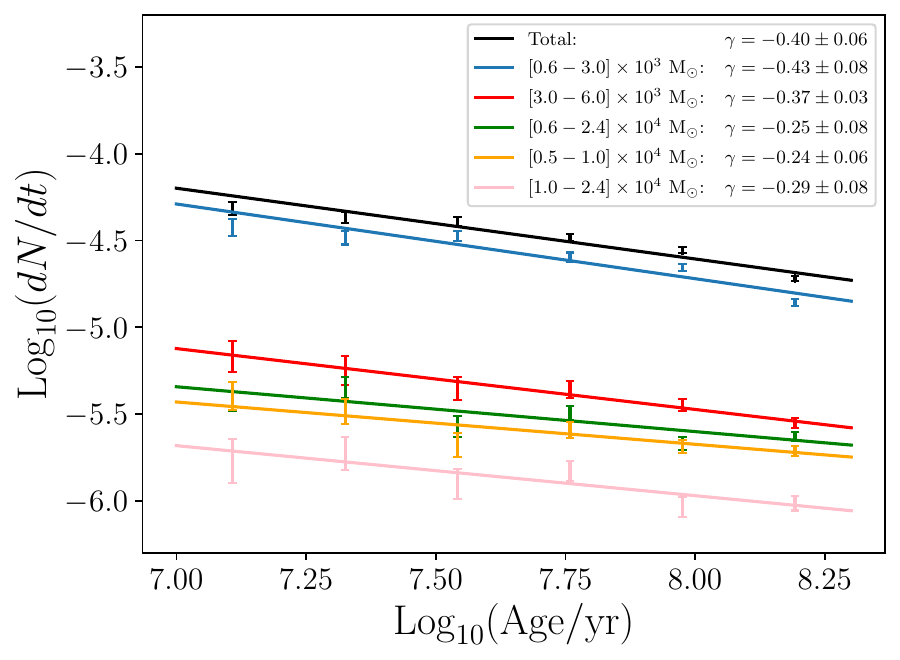}
 \caption{The age function for different mass intervals for the C$_{\rm{R}}$ clusters. The higher the cluster mass, the flatter the slope $\gamma$ becomes since the massive clusters are less sensitive to the galactic environment.}
 \label{fig:dNdt_age}
\end{figure}

For a cluster distribution with zero change, we would expect a flat distribution in age since our simulated clusters have a constant formation rate during the 200 Myr. All the age functions have been fitted to a single power-law, $dN/dt \propto t^{\gamma}$, in the age range from 10 to 200 Myr where a constant bin size in log$_{10}$ has been used. We exclude clusters with ages less than 10 Myr to be able to directly compare our study with the results of \citetalias{Messa2018}, where the age cut was made to avoid contamination of unbound clusters. The 10 Myr age cut is based on the fact that clusters older than 10 Myr have ages larger than their crossing time and can therefore be considered to be bound. The age functions can be seen in Figure \ref{fig:dNdt_age}, where we see the largest disruption for the least massive clusters with $\gamma = -0.43\pm 0.08$. As the cluster masses become greater, the slope of the age function becomes flatter which agrees well with the trend observed by \citetalias{Messa2018}. This is because the higher mass clusters are less affected by disruption caused by the galactic environment. We see a similar trend for the C$_{\rm{N}}$ clusters, but with a slightly more flattened slope in each mass interval, since these clusters have not had any interactions with GMCs. 

The mass range of $[0.5 - 1.0] \times 10^4$ \Mo (orange) shown in Figure \ref{fig:dNdt_age} can be directly compared to the observations of \citetalias{Messa2018}. We find a much flatter slope of $\gamma = -0.24\pm 0.06$ compared to the observed value of $\gamma = -0.39 \pm 0.04$. \citetalias{Messa2018} also calculated the age function for clusters in the mass range $[1.0 - 3.0] \times  10^4$ \Mo which we are almost able to cover, with our range being $[1.0 - 2.4] \times  10^4$ \Mo (pink). Here, we find an age function with $\gamma = -0.29 \pm 0.08$ which is in much better agreement with the observations of $\gamma -0.23\pm 0.07$. Our age function for the mass range $[0.5 - 1.0] \times 10^4$ \Mo shows a large discrepancy with the observations of \citetalias{Messa2018}. On the other hand, the mass range of $[1.0 - 3.0] \times 10^4$ \Mo matches well with the observations. A possible explanation is that the GMCs in our model are not destructive enough which would have the biggest impact on the lower mass clusters. This could be due to the high initial velocity dispersion in our model that the GMC are born with which, according to Eq. \ref{eq:dE}, will reduce the kinetic energy input of a cluster during a GMC encounter. 

To test whether the high velocity dispersion used is causing the discrepancy in the age function, we carried out $N$-body simulations of an isolated cluster, i.e. with no tides, that encounter a typical GMC in our model. A typical GMC encounter will have a GMC with maximum mass of $2 \times 10^6$ \Mo and an impact parameter of 40 pc with a relative velocity of $\sim 40$ \kms. Following an initial velocity distribution similar to that of the Milky Way of 7 \kmso \citep{Gustafsson2016, Jorgensen2020} will result in a GMC encounter with a relative velocity of typically $\sim 10$ \kms. We therefore ran several $N$-body simulations of a single GMC encounter with a cluster with an initial mass of 6000 \Mo and a typical half-mass radius of 2.5 pc. For each encounter, we varied the relative velocity between the cluster and GMC from 10 to 40 \kms. The starting position of the GMC was setup in such a way, that the point of closest approach would always be when the GMC had grown to its maximum mass. Each simulation lasted 50 Myr and the GMC was born 10 Myr after the start of the simulation to make sure that the cluster would be in virial equilibrium before interacting with the GMC. After 40 Myr the GMC died off and we continued the simulation for another 10 Myr in order to make sure that the stars have had enough time to escape the cluster. We classified all stars within a radius of 6.7 pc to be cluster members. 

We found that the GMC encounter with a relative velocity of 10 \kmso caused the cluster to be left with $\sim 56$ per cent of its stars, compared to a typical relative velocity in our model of $\sim 40$ \kms, where the cluster was left with $\sim 75$ per cent of its stars. We then scaled this difference with the mass difference between C$_{\rm{R}}$ and C$_{\rm{N}}$ clusters, for clusters where the C$_{\rm{R}}$ clusters had lost more mass than the C$_{\rm{N}}$ clusters. The mass loss was further scaled with the cluster initial half-mas radius, $r_{h,\rm{i}}$, and initial mass, $M_i$, following Eq. \ref{eq:dE} as $\propto r_{h,\rm{i}}^3 \, M_i^{-1}$. We were able to do this, since the mass difference between a C$_{\rm{R}}$ and C$_{\rm{N}}$ cluster is a direct result of the presence of the GMCs. Using this approach, we found an age function for the mass interval $[0.5 - 1.0] \times 10^4$ \Mo of $\gamma = -0.26\pm 0.06$ which is similar to our original result. Even if we double this effect we still see little difference, and not anything near the same age function which is observed. The reason for this is that the mass difference between the C$_{\rm{R}}$ and C$_{\rm{N}}$ clusters is only a small fraction of the total initial mass of the clusters. So even if the disruptiveness of the GMC encounters is doubled or tripled, the overall mass loss of the clusters is still dominated by the rest of the galactic tidal field.  %Furthermore, clusters which were original in our mass interval are replaced by clusters with masses above $10^4$ \Mo that now also lose more mass.    

Another reason for the discrepancy in the age function could be that higher mass clusters are born with larger initial half-mass radii than the less massive clusters and they do therefore not follow our ICRF given in Eq. \ref{eq:ICRF}. This is supported by \citet{Brown2021}, who found that clusters with ages between 1 and 10 Myr show a weak correlation between effective radius and mass as $R_{\rm{eff}} \propto M^{\beta_0}$, with $\beta_0 = 0.180\pm 0.028$. If we only consider clusters which are born with an initial half-mass radius above 2 pc, we get $\gamma = -0.44\pm 0.06$ which matches the observations. However, we then also find a $\gamma = -0.46\pm 0.12$ for the mass interval $[1.0 - 2.4] \times 10^4$ \Mo which is now a very poor result compared with the observations by \citetalias{Messa2018} for clusters in the mass interval $[1.0 - 3.0] \times 10^4$ \MO. It is therefore unlikely that the discrepancy stems from the ICRF. 

From our results and analysis, we are able to obtain the same results as \citetalias{Messa2018} in the mass interval $[1.0 - 2.4] \times 10^4$ \Mo, but not for the clusters in the mass $[0.5 - 1.0] \times 10^4$ \Mo. This is the lowest observed mass range by \citetalias{Messa2018} which also means that this interval has the highest likelihood of being incomplete. This is also addressed by \citetalias{Messa2018}: \lq Incompleteness at ages $\approx$200 Myr may start affecting the less massive sources ($\sim 5000$ \MO), which seem to have signifcantly more disruption\rq. If this is indeed the case, it explains why we are seeing such a big discrepancy between the observations of \citetalias{Messa2018} and our results for the age function in this mass interval.

\subsection{Environmental dependence}
Inspired by \citetalias{Messa2018b}, we investigated the evolution of the cluster populations by their galactocentric radius and whether they are located in a SA or IA region. We divided the clusters into 4 radial bins which contain roughly equal numbers of clusters in the range of 1.3 to 6.0 kpc. 

For the SA and IA regions we find that $\sim 46$ per cent of clusters are located in a IA region compared with \citetalias{Messa2018b} who found a value of $\sim 59$ per cent. Based on data of the PAWS \citep{Schinnerer2013}, \citetalias{Messa2018b} derived the average surface density of $\rm{H}_2$ for clusters in the SA and IA regions to be 55.3 and 16.5 \Mo $\rm{pc}^{-2}$, respectively. By evaluating the surface density of the gas disc and spiral arms, we find an average gas surface density, $\langle \Sigma_{\rm{gas}} \rangle$, of 60.2 \Mo $\rm{pc}^{-2}$ for clusters in the SA region which is in good agreement with observations. For the IA region, we find a value of 37.0 \Mo $\rm{pc}^{-2}$ which is significantly higher than observations. This discrepancy is likely related to how we define the SA and IA regions, which we based on the GMC distribution from the observations of \citet{Colombo2014}. If we increase the area of the SA region to include all clusters within 400 pc of the spiral locus (referred to as SA$_{400}$ and IA$_{400}$), we get a $\langle \Sigma_{\rm{gas}} \rangle$ of 17.8 \Mo $\rm{pc}^{-2}$ for the IA$_{400}$ region, while $\langle \Sigma_{\rm{gas}} \rangle$ stays unchanged in the SA$_{400}$ region. Increasing the area of the SA region also drops the fraction of GMCs in the IA$_{400}$ region down to $\sim 20$ per cent, indicating that our fraction of 10 per cent of GMC, which are randomly born in the disc, might be too low in our galactic model. In our further analysis, we consider both definintions of the SA and IA regions which are defined by the distance to the spiral locus of 250 and 400 pc. The area and the number of C$_{\rm{R}}$ clusters for each region can be seen in Table \ref{table:region}. 

\begin{table}
 \begin{tabular*}{\columnwidth}{@{\extracolsep{\fill}} lccc}
  Region &  Area & $N_c$ & $N_c$\\
  \hline
  
    & [kpc] \Tstrut\Bstrut &  $M \ge 600$\MO   & $M \ge 5000$\MO \\
  Bin 1       &1.3 - 2.0                       &  619 & 183\\
  Bin 2       &2.0 - 2.8                       &  719 & 185\\
  Bin 3       &2.8 - 4.0                       &  600 & 162\\
  Bin 4       &4.0 - 6.0                       &  647 & 167\\
  SA$_{250}$  &$d_{\mathrm{locus}} \le 0.25$   & 1418 & 397\\
  IA$_{250}$  &$d_{\mathrm{locus}} > 0.25$     & 1167 & 300\\
  SA$_{400}$  &$d_{\mathrm{locus}} \le 0.40$   & 1911 & 528\\
  IA$_{400}$  &$d_{\mathrm{locus}} > 0.40$     &  674 & 169\\
  Bin$_{\rm{M}}$ 1  &1.30 - 3.17               & 1558 & 429\\
  Bin$_{\rm{M}}$ 2  &3.17 - 4.54               &  598 & 160\\
  \hline
 \end{tabular*}
 \caption{The area defined for each region of M51. The number of C$_{\rm{R}}$ clusters in each region, $N_c$, is shown for a mass cut of 600 and 5000 \MO.}
 \label{table:region}
\end{table}

\subsubsection{Mass function}
\begin{figure}
 \includegraphics[width=0.8\columnwidth]{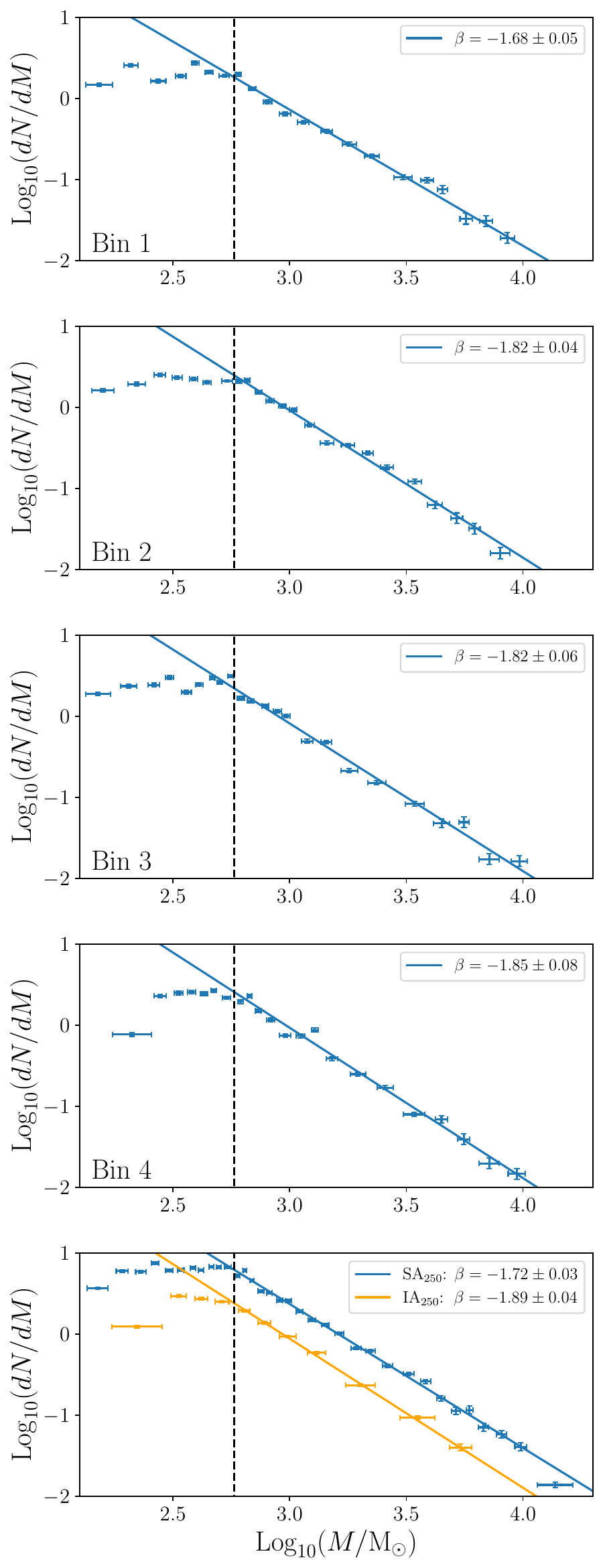}
 \caption{The mass functions of different regions that can be seen in Table \ref{table:BIN_FITS}. The mass functions have each been fitted to a single power-law of slope $\beta$.}
 \label{fig:Bin_MF}
\end{figure}

Each mass function is fitted to a single power-law where the best fits are listed in Table \ref{table:BIN_FITS}, and the mass functions for the C$_{\rm{R}}$ clusters can be seen in Figure \ref{fig:Bin_MF}. As the galactocentric radius increases for the C$_{\rm{R}}$ clusters, the slope of the mass function steepens. Clusters closer to the galactic centre are more exposed to stronger tidal forces and GMC encounters which mostly affect the lower mass clusters. There is a similar trend for the C$_{\rm{N}}$  clusters, however, for Bin 3 the mass function decreases significantly, before increasing again for Bin 4. Besides this, we see no significant difference between the C$_{\rm{R}}$ and C$_{\rm{N}}$ clusters. Adopting a similar mass cut as \citetalias{Messa2018b} of 5000 \MO, we see $\beta \sim -2$ for all bins which was also what observations showed when fitted to a single power-law. We also included the two first bin ranges used in \citetalias{Messa2018b} which are labelled as Bin$_{\rm{M}}$. For Bin$_{\rm{M}}$ 1 and 2, we find slopes of $-2.10\pm 0.14$ and $-2.03\pm 0.17$ in good agreement with observations of $-2.06$ and $-2.18$, respectively.

The slope of the mass function for the SA$_{250}$ region, $\beta = -1.72\pm 0.03$, is similar to the slope of Bin 1, $\beta = -1.68\pm 0.05$. This is largely to do with the fact that the spiral arms have curled up in the centre which means that much of the area in the centre $R \sim [1.3 - 1.8]$ kpc is classified as part of the SA region. This is also seen in \citetalias{Messa2018b} where Bin$_{\rm{M}}$ 1 has a slope of $-1.77$ and their SA region has a slope of $-1.76$ for their truncated fits. We find a slope for the IA$_{250}$ region to be $\beta = -1.89\pm 0.04$ which is significantly steeper than for the SA$_{250}$ region. The same trend was found in the observations of \citetalias{Messa2018b}. For the SA$_{400}$ and IA$_{400}$, we find $\beta$ $-1.78\pm 0.03$ and $-1.84\pm 0.03$, respectively. We see that their $\beta$ starts to converge together, since the SA$_{400}$ region dominates most of the cluster popluation. Applying the same mass cut as in \citetalias{Messa2018b}, we find $\beta = -2.08\pm 0.15$ for the SA$_{250}$ region compared to observation of $-1.90\pm 0.03$. In the IA$_{250}$ region we find $-1.89\pm 0.21$ compared to observations of $-2.08\pm 0.04$. At first glance, it seems that there is more cluster disruption in the IA region, however, the uncertainty of the slope is high because of the low number of clusters in the region.

\subsubsection{Age function}
\begin{figure}
 \includegraphics[width=1.0\columnwidth]{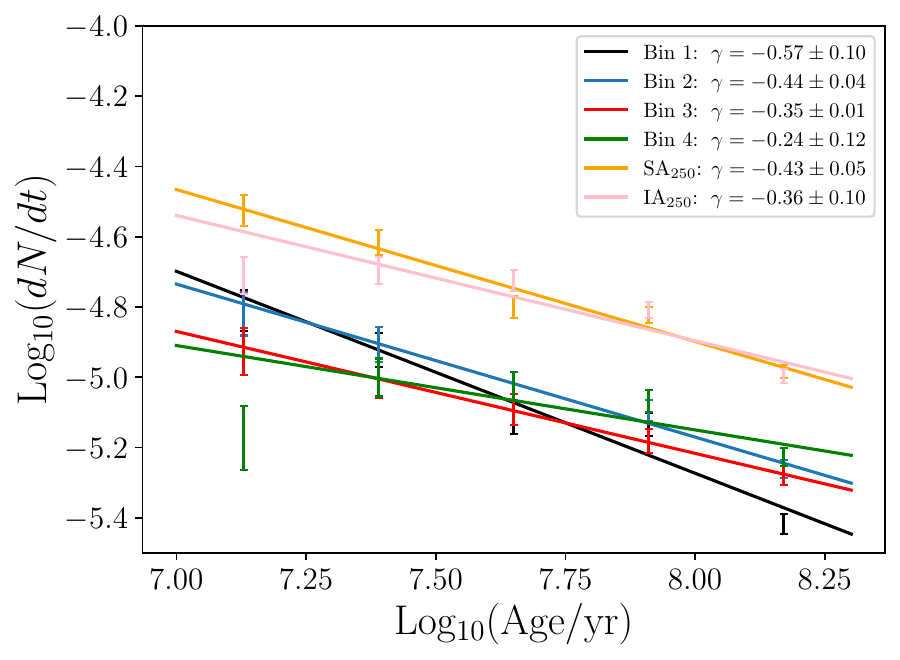}
 \caption{The age functions for each region described in Table \ref{table:BIN_FITS}.}
 \label{fig:Bin_age}
\end{figure}

The age function for the C$_{\rm{R}}$ clusters in each region can be seen in Figure \ref{fig:Bin_age}, where the fits are listed in Table \ref{table:BIN_FITS}. We tried several bin sizes for constructing the age function and found that the slope for each region was within the estimated $1\sigma$ uncertainty, and thus our results are not dependent on the bin size. We see stronger disruption of clusters that are located close to the galactic centre, with Bin 1 having the steepest slope of $\gamma = -0.57\pm 0.10$. As the distance from the galactic centre inceases, the slope flattens with Bin 2 and Bin 3 having slopes of $-0.44 \pm 0.04$ and $-0.35 \pm 0.01$, respectively. Bin 4 has the flattest slope with $\gamma = -0.24\pm 0.12$. The C$_{\rm{N}}$ clusters follow the same pattern, albeit with a slightly more flattened slope for each bin. Applying the mass cut of \citetalias{Messa2018b}, we see a clear decrease in the disruption of clusters, with Bin 1 now having a slope of $-0.33\pm 0.15$. There is less difference between Bin 1, 2 and 3 which indicates that the most massive clusters are less affected by their radial location. It is only Bin 4 that differs with $\gamma = -0.10\pm 0.02$, where the disruption of clusters has dropped significantly. There is no clear difference between the C$_{\rm{R}}$ and C$_{\rm{N}}$ age functions which is reasonable, since the most massive clusters are less affected by GMC encounters. For Bin$_{\rm{M}}$ 1, we find $\gamma = -0.31\pm 0.01$ which is significantly lower than observation of $-0.50\pm 0.9$. Bin$_{\rm{M}}$ 2 is a much better match with $\gamma = -0.30\pm 0.11$ compared to observations of $-0.38 \pm 0.07$. 

We find the biggest difference in disruption between the SA$_{400}$ and IA$_{400}$ regions which have $\gamma = -0.45\pm 0.03$ and $\gamma = -0.25\pm 0.15$, respectively. Considering the mass cut of \citetalias{Messa2018b}, we find less disruption with SA$_{400}$: $-0.31\pm 0.03$ and IA$_{400}$: $-0.22\pm 0.01$. In case of the IA region, \citetalias{Messa2018b} found $\gamma = -0.15\pm 0.03$ which matches well with our results for IA$_{400}$, especially for the C$_{\rm{N}}$ clusters where $\gamma = -0.19\pm 0.01$. For the SA region, \citetalias{Messa2018b} found $\gamma = -0.73\pm 0.07$ compared to our most disruptive SA which is SA$_{400}$ with $\gamma = -0.31\pm 0.03$. This is a large discrepancy, and it should be noted that the derived age function for the SA and IA regions for the mass cut of 5000 \Mo is very sensitive to how the different regions are defined. This is evident for the SA$_{250}$ and IA$_{250}$ regions, where we measure more cluster disruption in the IA$_{250}$ region. Even so, if we consider our whole cluster population, without any mass cut, our most disruptive $\gamma$ for the SA region is only $-0.45\pm 0.03$.

\begin{table*}
 \begin{tabular*}{\textwidth}{@{\extracolsep{\fill}} lcccccccc}
  %\hline
  Region & C$_{\rm{R}}$ & C$_{\rm{N}}$ & C$_{\rm{R}}$ & C$_{\rm{N}}$ & C$_{\rm{R}}$ & C$_{\rm{N}}$ & C$_{\rm{R}}$ & C$_{\rm{N}}$ \\
  \hline
    &    \multicolumn{4}{c}{$\beta$}  & \multicolumn{4}{c}{$\gamma$} \\
  \cmidrule{2-5}\cmidrule{6-9} 
    & \multicolumn{2}{c}{$M \ge 600$\MO}  & \multicolumn{2}{c}{$M \ge 5000$\MO} & \multicolumn{2}{c}{$M \ge 600$\MO}  & \multicolumn{2}{c}{$M \ge 5000$\MO} \\
  \cmidrule{2-3}\cmidrule{4-5}\cmidrule{6-7}\cmidrule{8-9} 	
  Bin 1       &       $-1.68\pm 0.05$  & $-1.70\pm 0.06$ & $-2.32\pm 0.21$ & $-2.18\pm 0.21$ &  $-0.57\pm 0.10$  & $-0.51\pm 0.09$ & $-0.33\pm 0.15$ & $-0.28\pm 0.15$\\
  Bin 2       &       $-1.82\pm 0.04$  & $-1.84\pm 0.06$ & $-2.02\pm 0.22$ & $-2.05\pm 0.20$ &  $-0.44\pm 0.04$  & $-0.41\pm 0.03$ & $-0.39\pm 0.22$ & $-0.33\pm 0.22$\\
  Bin 3       &       $-1.82\pm 0.06$  & $-1.92\pm 0.09$ & $-1.99\pm 0.28$ & $-2.15\pm 0.26$ &  $-0.35\pm 0.01$  & $-0.29\pm 0.02$ & $-0.36\pm 0.13$ & $-0.32\pm 0.12$\\
  Bin 4       &       $-1.85\pm 0.08$  & $-1.82\pm 0.06$ & $-1.95\pm 0.28$ & $-2.07\pm 0.26$ &  $-0.24\pm 0.12$  & $-0.21\pm 0.12$ & $-0.10\pm 0.02$ & $-0.07\pm 0.03$\\
  SA$_{250}$  &       $-1.72\pm 0.03$  & $-1.80\pm 0.04$ & $-2.08\pm 0.15$ & $-2.10\pm 0.18$ &  $-0.43\pm 0.05$  & $-0.37\pm 0.04$ & $-0.24\pm 0.04$ & $-0.19\pm 0.05$\\
  IA$_{250}$  &       $-1.89\pm 0.04$  & $-1.78\pm 0.05$ & $-1.89\pm 0.21$ & $-1.94\pm 0.16$ &  $-0.36\pm 0.10$  & $-0.33\pm 0.10$ & $-0.34\pm 0.01$ & $-0.31\pm 0.01$\\
  SA$_{400}$  &       $-1.78\pm 0.03$  & $-1.79\pm 0.05$ & $-2.02\pm 0.13$ & $-1.94\pm 0.15$ &  $-0.45\pm 0.03$  & $-0.40\pm 0.02$ & $-0.31\pm 0.03$ & $-0.26\pm 0.03$\\
  IA$_{400}$  &       $-1.84\pm 0.03$  & $-1.78\pm 0.04$ & $-1.79\pm 0.37$ & $-1.95\pm 0.33$ &  $-0.25\pm 0.15$  & $-0.23\pm 0.16$ & $-0.22\pm 0.01$ & $-0.19\pm 0.01$\\
  Bin$_{\rm{M}}$ 1  & $-1.73\pm 0.04$  & $-1.78\pm 0.04$ & $-2.10\pm 0.14$ & $-2.14\pm 0.13$ &  $-0.48\pm 0.09$  & $-0.22\pm 0.06$ & $-0.31\pm 0.01$ & $-0.26\pm 0.01$\\
  Bin$_{\rm{M}}$ 2  & $-1.86\pm 0.04$  & $-1.91\pm 0.08$ & $-2.03\pm 0.17$ & $-2.15\pm 0.21$ &  $-0.29\pm 0.09$  & $-0.25\pm 0.01$ & $-0.30\pm 0.11$ & $-0.26\pm 0.11$\\  
  \hline
 \end{tabular*}
 \caption{Results of the different regions for the whole cluster population, and when a mass cut of 5000 \Mo has been applied. Every region has been fitted to a single power-law, with slope $\beta$ for the mass function and an age function with slope $\gamma$.}
 \label{table:BIN_FITS}
\end{table*}

We underestimate the disruption of clusters in the innermost part of M51 compared to observations. The same is true for the spiral arm region where we see the biggest difference. \citetalias{Messa2018b} mentioned that there seems to be some cluster incompleteness in their SA region which could help explain the discrepancy between their observations and our results. Spiral arms are among the regions with the brighest background and largest crowding sources. Older and less massive clusters can therefore more easily be missed by observations which would also increase the derived disruption in the region up. Furthermore, the discrepancies could also be driven by stochastic effects, where the presence or absence of individual red supergiants makes it difficult to reliably determine the ages of low-mass clusters with ages $\le 50$ Myr. Investigating the disruption of clusters in M101, as a function of environment in the form of gas surface density, \citet{Linden2022} found good agreement with the observations by \citetalias{Messa2018b} if Bin$_{\rm{M}}$ 1 and the SA region were omitted. This further suggests that the disruption in these regions might be overestimated.

\section{The impact of GMC encounters}
\label{sec:Results2}
Even though there are little detectable difference between C$_{\rm{R}}$ and C$_{\rm{N}}$ clusters in the overall shape of the mass functions, the individual clusters are still different. In this section, we are going to investigate these differences and the possible causes of them.

\subsection{Cluster sizes and mass loss}
\begin{figure*}
 \includegraphics[width=1.0\textwidth]{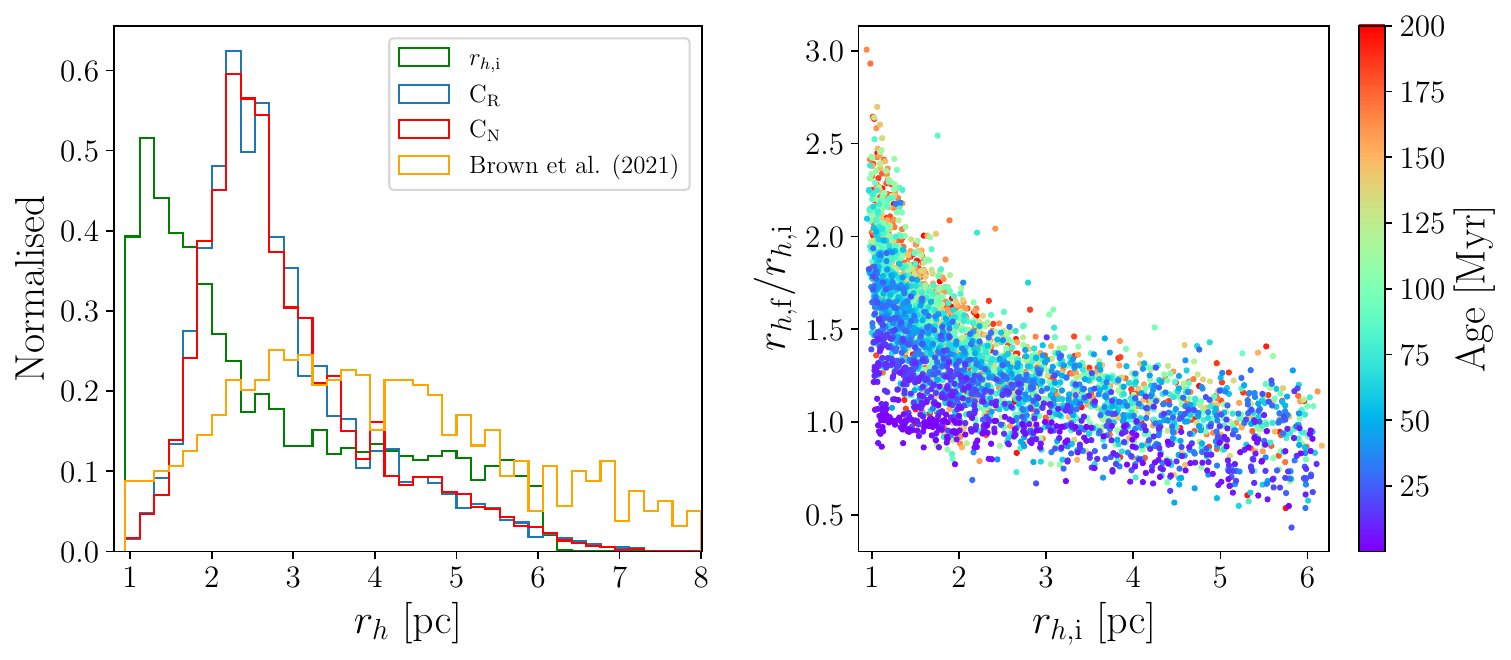}
 \caption{\textit{Left:} The distribution of half-mass radii for the initial clusters, together with the final half-mass radii for the C$_{\rm{R}}$ and C$_{\rm{N}}$ clusters, and the M51 cluster radii from the cluster catalogue of \citet{Brown2021}. Only clusters in the mass range $[600 - 24000]$ \Mo are shown in order to make a direct comparison. \textit{Right:} The ratio between the final and initial half-mass radius for the C$_{\rm{R}}$ clusters as a function of initial half-mass radius.}
 \label{fig:rh}
\end{figure*}

The initial half-mass radius of each cluster, $r_{h,\rm{i}}$, is given by the ICRF in Eq. \ref{eq:ICRF} and can be seen to the left in Figure \ref{fig:rh}, together with the final half-mass radii, $r_{h,\rm{f}}$, of the surviving C$_{\rm{R}}$ and C$_{\rm{N}}$ clusters. Only clusters in the mass range $[600 - 24000]$ \Mo are shown in order to compare with the M51 cluster catalogue of \citet{Brown2021}. The effective radii from \citet{Brown2021} have been converted to half-mass radii by the relation $r_h = 1.6 R_{\rm{eff}}$ from \citet{Hurley2010}. The distributions of the C$_{\rm{R}}$ and C$_{\rm{N}}$ clusters are quite similar and both are shifted towards higher radii compared to their initial birth value. The similarity of the C$_{\rm{R}}$ and C$_{\rm{N}}$ clusters indicates that the GMCs have little effect on the distribution of the cluster radii. The radial distribution of our simulated clusters shows a clear peak around 2.5 pc, whereas the clusters of \citet{Brown2021} have a much more even distribution. This difference is likely because we restrict the membership of our clusters based on the energy of each individual star and the tidal radius, which reduces the radial extent of our clusters. The discrepancy with our cluster radii and the catalogue of \citet{Brown2021} could also indicate that our ICRF is underestimating the cluster sizes. However, our ICRF is in agreement with the \citet{Brown2021} catalogue for clusters with ages less than 10 Myr. Furthermore, the age function analysis in section 4.2, showed that increasing the radii of the cluster distribution leads to a too high disruption of the higher mass clusters compared to observations. It is therefore difficult to justify an increase in the initial radii of the cluster distribution in order to match our simulations to the observations. 

To the right in Figure \ref{fig:rh}, the ratio between the final half-mass radius $r_{h,\rm{f}}$ and $r_{h,\rm{i}}$ can be seen as a function of $r_{h,\rm{i}}$ for the C$_{\rm{R}}$ clusters. Clusters with young ages have not had enough time to expand and therefore have $r_{h,\rm{f}}/r_{h,\rm{i}} \sim 1$. Some of the clusters with high $r_{h,\rm{i}}$ even contract which is likely because they are already filling their tidal radius and are in the process of losing stars. As the age of the clusters increases, $r_{h,\rm{f}}/r_{h,\rm{i}}$ gets larger and goes beyond a value of 2 for the clusters which start off with small radii. The oldest clusters seem to be creating an arc which is caused by the fact that they have had more time to evolve via two-body relaxation and to be affected by the galactic tidal field. \citet{Chandar2016} observed that clusters in M51 more massive than $\sim 3 \times 10^4$ \Mo increase in half-light radii by a factor of $\sim 3-4$ over the first few hundred Myr. We find that the average increase in half-mass radii for both C$_{\rm{R}}$ and C$_{\rm{N}}$ clusters is $\sim 1.4$, with 6 per cent of the clusters having an expansion factor higher than 2. Since we are operating with less massive clusters than \citet{Chandar2016}, our clusters cannot expand as much due to the stripping of stars by the galactic tidal field which could be the reason for this discrepancy. We also compared our cluster radii to the catalogue of \citet{Brown2021} in the mass range $[0.6-2.4] \times 10^4$ \Mo for clusters with ages in the ranges [0,10] Myr and [150,200] Myr. We find a median increase of $\sim 1.2$ compared to  \citet{Brown2021} where a value of $\sim 1.5$ is observed.      

\begin{figure*}
\includegraphics[width=1.0\textwidth]{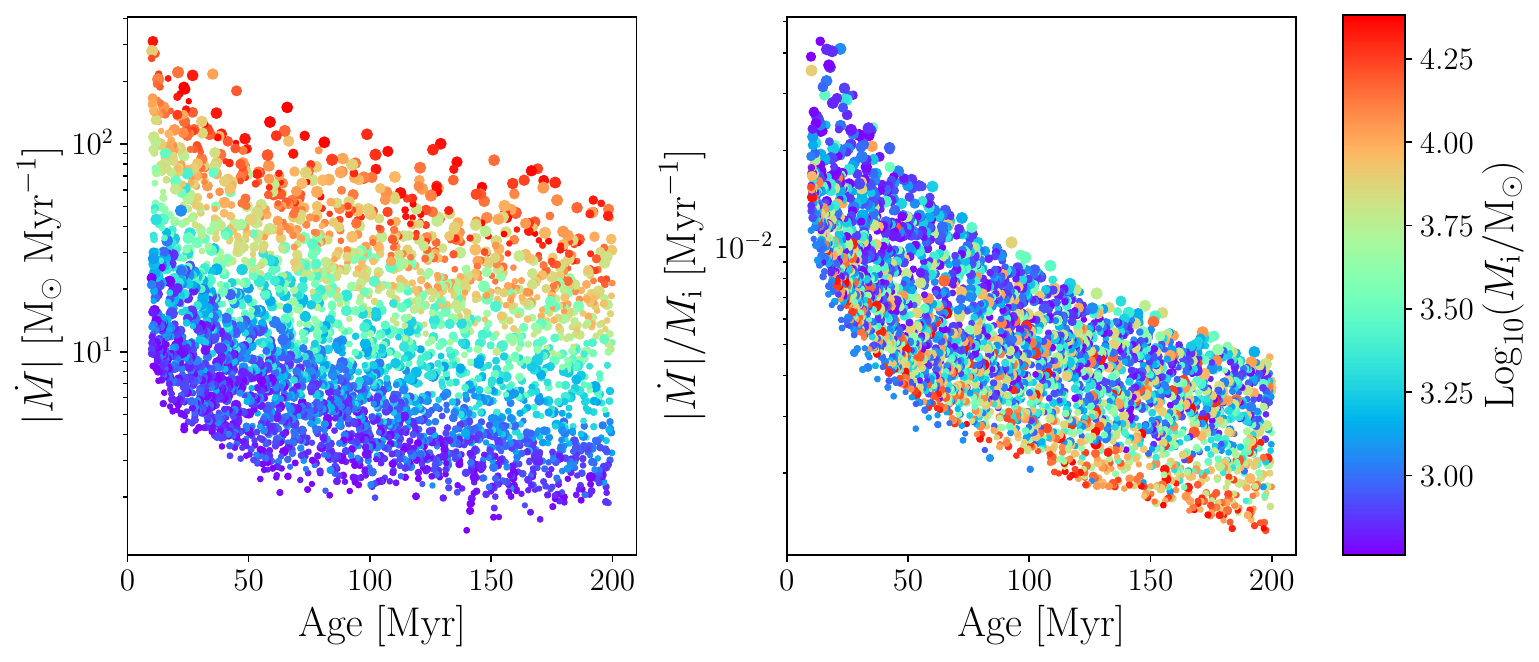}
\caption{\textit{Left:} The average C$_{\rm{R}}$ cluster mass loss rate as a function of age where the size of each data point is scaled with the initial half-mass radius. Clusters younger than 10 Myr have been excluded to remove the most extreme effects of stellar feedback. \textit{Right:} The average fractional mass loss rate as a function of age.}
\label{fig:dMdt_age}
\end{figure*}

The average mass loss rate for each of the surviving C$_{\rm{R}}$ clusters can be seen to the left in Figure \ref{fig:dMdt_age} as a function of age. Here, the size of the data points for each cluster has been scaled relative to the initial cluster half-mass radius. We exclude clusters younger than 10 Myr in order to remove the most extreme effects of mass loss caused by stellar evolution for the most massive stars. We see a large spread between the mass loss rate experienced for each cluster, and a general trend where the mass loss rate decreases as the clusters gets older. This can be attributed to stellar evolution. There is a clear dependence on the initial half-mass radius where the more compact clusters lose much less mass. It is somewhat surprising that the most massive clusters experience the largest amount of mass loss. At a first glance, one might naively think that the less massive clusters would be more sensitive to the galactic field and therefore have a higher mass loss rate compared to the massive clusters. This is indeed the case, but the more massive clusters have much more mass which is lost via stellar feedback and they therefore dominate in the mass loss rate. This brings us to the right of Figure \ref{fig:dMdt_age}, where instead the average fractional mass loss rate of the clusters are shown. This gives a much better picture of the dynamics of the clusters where the low-mass clusters dominate the top of the fractional mass loss rate. The clusters which have the largest frational mass loss rate are those with low mass and high initial $r_h$. Up to around 100 Myr, there are several lower mass clusters in the bottom of the plot which forms an arch which then disappears due to an increase in mass loss. This increase is likely caused by the clusters filling their tidal radius and starting to lose stars. For the C$_{\rm{N}}$ clusters, we see a similar trend with no clear distinction from the C$_{\rm{R}}$ clusters.

\subsection{Cluster destruction}
\begin{figure}
 \includegraphics[width=\columnwidth]{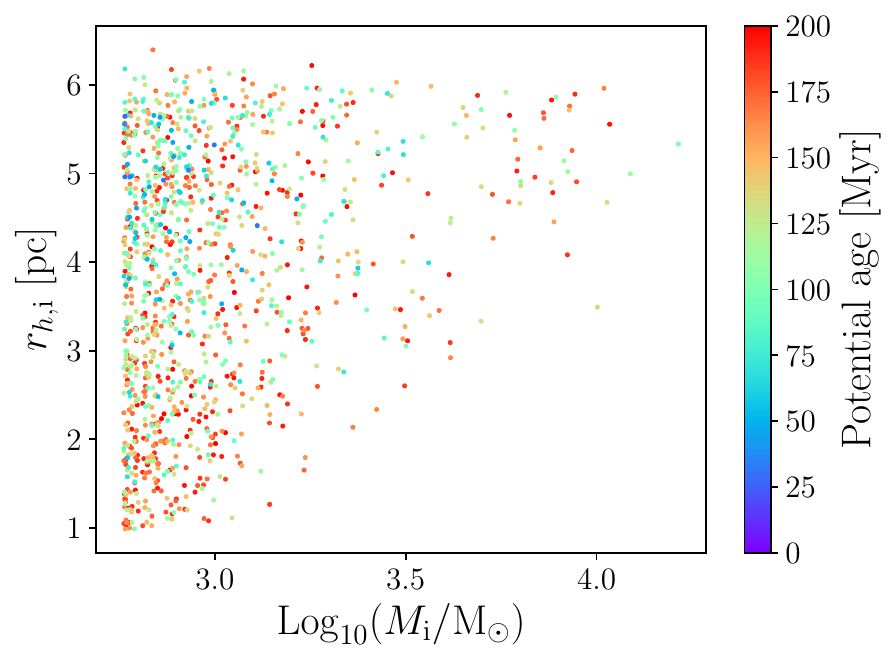}
 \caption{The destroyed C$_{\rm{R}}$ clusters shown with their initial masses and half-mass radii. They are colour coded according to the potential age they would have reached if they had not been destroyed.}
 \label{fig:Destroyed}
\end{figure}

We define a cluster to be destroyed if it has less than 200 stars. The destroyed C$_{\rm{R}}$ clusters are shown in Figure \ref{fig:Destroyed}, with their initial masses and half-mass radii. They have been colour coded in accordance with the age that they could potentially have reached had they not been destroyed. Low-mass clusters with high $r_{h,\rm{i}}$ are most easily destroyed and the more compact the low-mass clusters are, the longer it takes for them to be destroyed. The only high-mass clusters that are destroyed have high $r_{h,\rm{i}}$ and potential ages above $\sim 100$ Myr. We see the same pattern for the C$_{\rm{N}}$ clusters, but with 194 fewer destroyed clusters.     

\begin{figure}
 \includegraphics[width=\columnwidth]{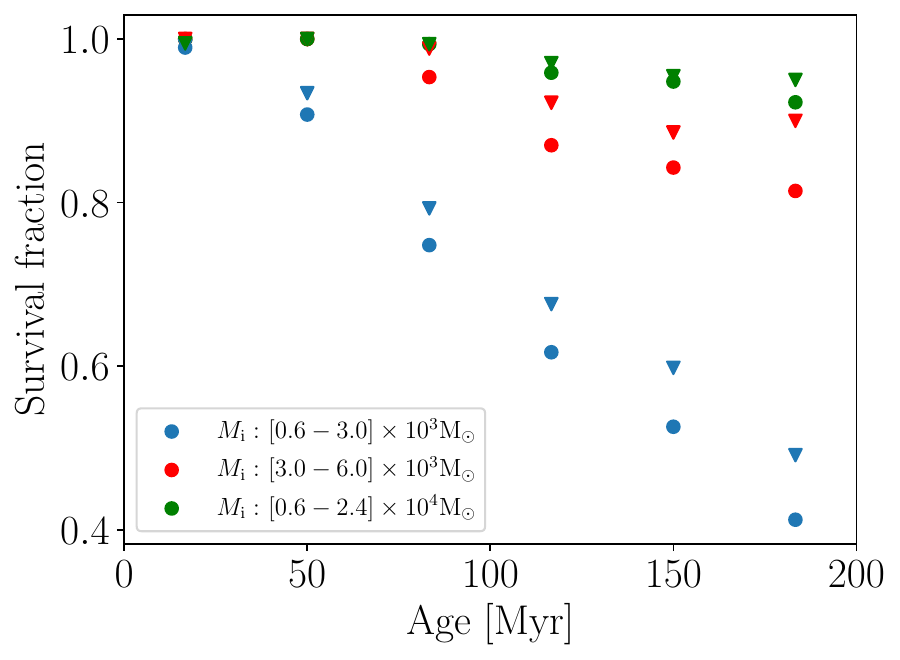}
 \caption{The fraction of clusters that survive as a function of their age where the clusters are divided into different initial mass ranges. The C$_{\rm{R}}$ and C$_{\rm{N}}$ clusters are marked as circles and triangles, respectively.}
 \label{fig:Survival}
\end{figure}

Figure \ref{fig:Survival} shows the fraction of clusters that survive during our simulations and are divided into three different mass intervals, where C$_{\rm{R}}$ and C$_{\rm{N}}$ clusters are represented by circles and triangles, respectively. For the more massive clusters (green), GMCs have little impact on the destruction of the clusters and we do not see a significant difference between the C$_{\rm{R}}$ and C$_{\rm{N}}$ clusters before the clusters reach an age of $\sim 180$ Myr. The importance of the GMC encounters becomes more pronounced as the initial cluster mass decreases (red), and a difference can be seen at an age of $\sim 80$ Myr. For the lowest mass clusters (blue), the difference is evident already after an age of $\sim 50$ Myr where 10 per cent of the clusters have been destroyed. For the least massive clusters, only $\sim 40$ per cent of them have survived at the age of 200 Myr whereas over 90 per cent of the most massive clusters are still surviving. For the oldest clusters with masses below 6000 \MO, we see a difference of $\sim$ 8 per cent points in the survival fraction between the C$_{\rm{R}}$ and C$_{\rm{N}}$ clusters.   
    
\subsection{Cluster groups}
\begin{figure}
\includegraphics[width=\columnwidth]{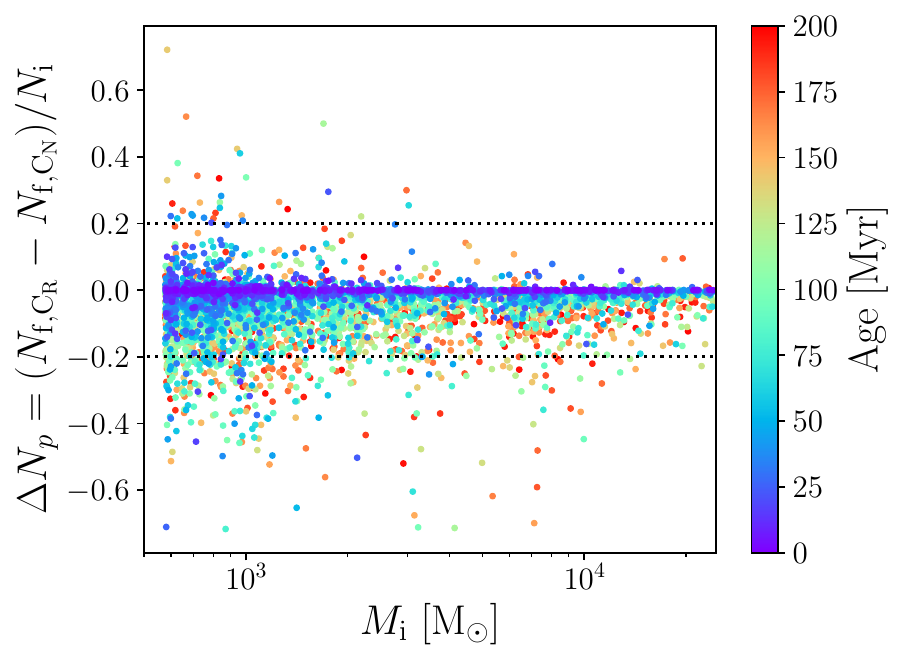}
\caption{$\Delta N_p$, the fractional difference between the final number of stars in the C$_{\rm{R}}$ and C$_{\rm{N}}$ clusters divided by the initial number of stars in the cluster, $N_{\rm{i}}$, as a function of initial cluster mass. $M_{\rm{i}}$.}
\label{fig:dNp}
\end{figure}

We now make a direct comparison between the number of stars left in each individual C$_{\rm{R}}$ and C$_{\rm{N}}$ cluster. We defined the parameter $\Delta N_p = (N_{\rm{f},\rm{C}_{\rm{R}}} - N_{\rm{f},\rm{C}_{\rm{N}}})/N_{\rm{i}}$ which is the fractional difference between the final number of stars $N_{\rm{f}}$ for the C$_{\rm{R}}$ and C$_{\rm{N}}$ clusters, where $N_{\rm{i}}$ is the initial number of stars for each cluster. $\Delta N_p$ can be seen in Figure \ref{fig:dNp} as a function of the initial mass of each cluster. For most of the clusters, we see that it is the C$_{\rm{R}}$ clusters which are left with the fewest stars at the end of the simulations. For young clusters there is only a spread of a few per cent in $\Delta N_p$, since these clusters have not had time to evolve and be affected by the galactic tides. As the clusters get older there is a stronger discrepancy between C$_{\rm{R}}$ and C$_{\rm{N}}$ clusters, and we see that the majority of the clusters have a decrease in $\Delta N_p$. This corresponds well with what we would expect, i.e. that GMCs have a destructive effect on stellar clusters. There is, however, a minority of clusters which end up having less stars when they are not affected by GMCs, which is quite surprising! To explore the differences in $\Delta N_p$ between the clusters, we divided them into three different groups: 
\begin{itemize}
\item[] P$_{\rm{C}}$: Predicted clusters with $\Delta N_p$ < -0.2.
\newline
\item[] U$_{\rm{C}}$: Unpredicted clusters with $\Delta N_p$ > 0.2.
\newline
\item[] N$_{\rm{C}}$: Neutral clusters with $|\Delta N_p|$ < 0.2.
\end{itemize} 
The threshold between the different cluster groups is depicted in Figure \ref{fig:dNp} as horizontal dotted lines. Clusters P$_{\rm{C}}$ and U$_{\rm{C}}$ represents the biggest differences in $\Delta N_p$, and C$_{\rm{N}}$ represents clusters which show little differences in $\Delta N_p$.

\begin{figure}
 \includegraphics[width=\columnwidth]{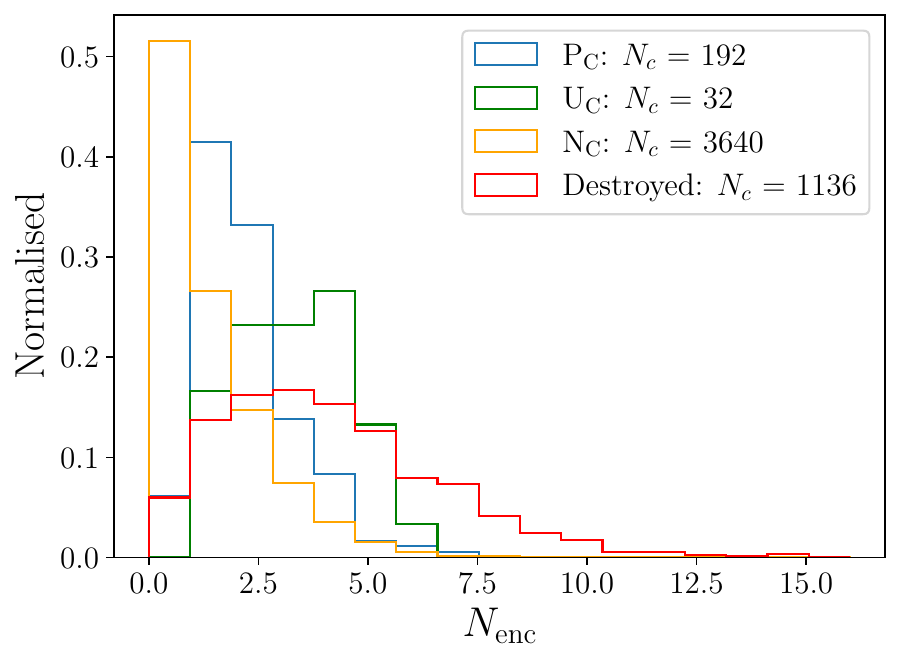}
 \caption{ The distribution of number of GMC encounters per cluster for the different cluster groups.}
 \label{fig:Nenc}
\end{figure}

The normalised distribution of the three cluster groups can be seen in Figure \ref{fig:Nenc} as a function of the number of strong GMC encounters they experience in \textsc{nbody6tt}. The destroyed C$_{\rm{R}}$ clusters have also been included together with the number of clusters in each group, $N_c$. As expected, over half of the N$_{\rm{C}}$ clusters do not experience any GMC encounters which makes the tidal evolution experienced by the C$_{\rm{R}}$ and C$_{\rm{N}}$ clusters very similar. The clusters that are destroyed experience the most GMC encounters. Out of these, 5 per cent do not experience any GMC encounters in \textsc{nbody6tt}, but are destroyed by the rest of the galactic tidal environment which also includes GMC encounters that could be approximated into the tidal tensor. The majority of P$_{\rm{C}}$ clusters experience at least one GMC encounter during their lifetime, but only a small fraction have more than five GMC encounters. U$_{\rm{C}}$ clusters all have one or more GMC encounters and a larger fraction of clusters which experiences several encounters compared to the distribution of P$_{\rm{C}}$ clusters. On average, U$_{\rm{C}}$ clusters experience 3.1 GMC encounters, whereas it is only 1.9 for P$_{\rm{C}}$ clusters. 

Because of the low number of U$_{\rm{C}}$ clusters, we were able to manually investigate all 32 clusters in detail. For 7 of the U$_{\rm{C}}$ clusters, we find that they have large fractions of their stars that have postive energies within the tidal radius which are in the process of escaping the clusters. They will therefore be destroyed in a couple of Myr and we are just "lucky" enough to catch them before they do so. There is therefore nothing surprising about these clusters. We find 10 U$_{\rm{C}}$ clusters that have a fraction of stars with positive energies that are also in the process of leaving the cluster. Given a few Myr and these U$_{\rm{C}}$ clusters will probably have $|\Delta N_p| <0.2$, i.e. the C$_{\rm{R}}$ and C$_{\rm{N}}$ clusters will look similar. For these clusters, it seems that the GMC encounters are increasing the kinetic energy of the stars, but the compressive tides are delaying the escape of these stars, and this is why we are seeing a difference between the C$_{\rm{R}}$ and C$_{\rm{N}}$ clusters. The rest 15 $U_P$ clusters do in fact represent a true difference between the C$_{\rm{R}}$ and C$_{\rm{N}}$ clusters and it is not just a case of us finding clusters which are in the process of being destroyed or losing a large fraction of their stars. For these 15 $U_P$ clusters, we see different evolutionary trends which have an impact on the difference between C$_{\rm{R}}$ and C$_{\rm{N}}$ clusters. We see examples of C$_{\rm{R}}$ clusters which have one or two GMC encounters at a very young age, which in return determines the rest of the cluster evolution. Other clusters with $|z|$-components above 100 pc experience disc shocks, but during some of these disc shocks they also experience GMC encounters, especially if they are passing through a spiral arm, causing the C$_{\rm{R}}$ clusters to lose less stars than the C$_{\rm{N}}$ clusters. Finally, we have clusters which spend their life in the galactic disc, i.e. $|z|< 100$ pc, which have many interactions with the spiral arms while also having several GMC encounters. 

In order to explain the existence of these 15 U$_{\rm{C}}$ clusters, we propose that several of the GMCs encounters do in fact protect the C$_{\rm{R}}$ clusters which cause them to lose less stars than the C$_{\rm{N}}$ clusters. This is due to the GMC encounters creating tides that cancel out the tides created by the rest of the galactic field. However, there are only 15 U$_{\rm{C}}$ clusters which makes this a rare occurrence.    

\subsection{Adiabatic galactic tides}
\begin{figure}
 \includegraphics[width=\columnwidth]{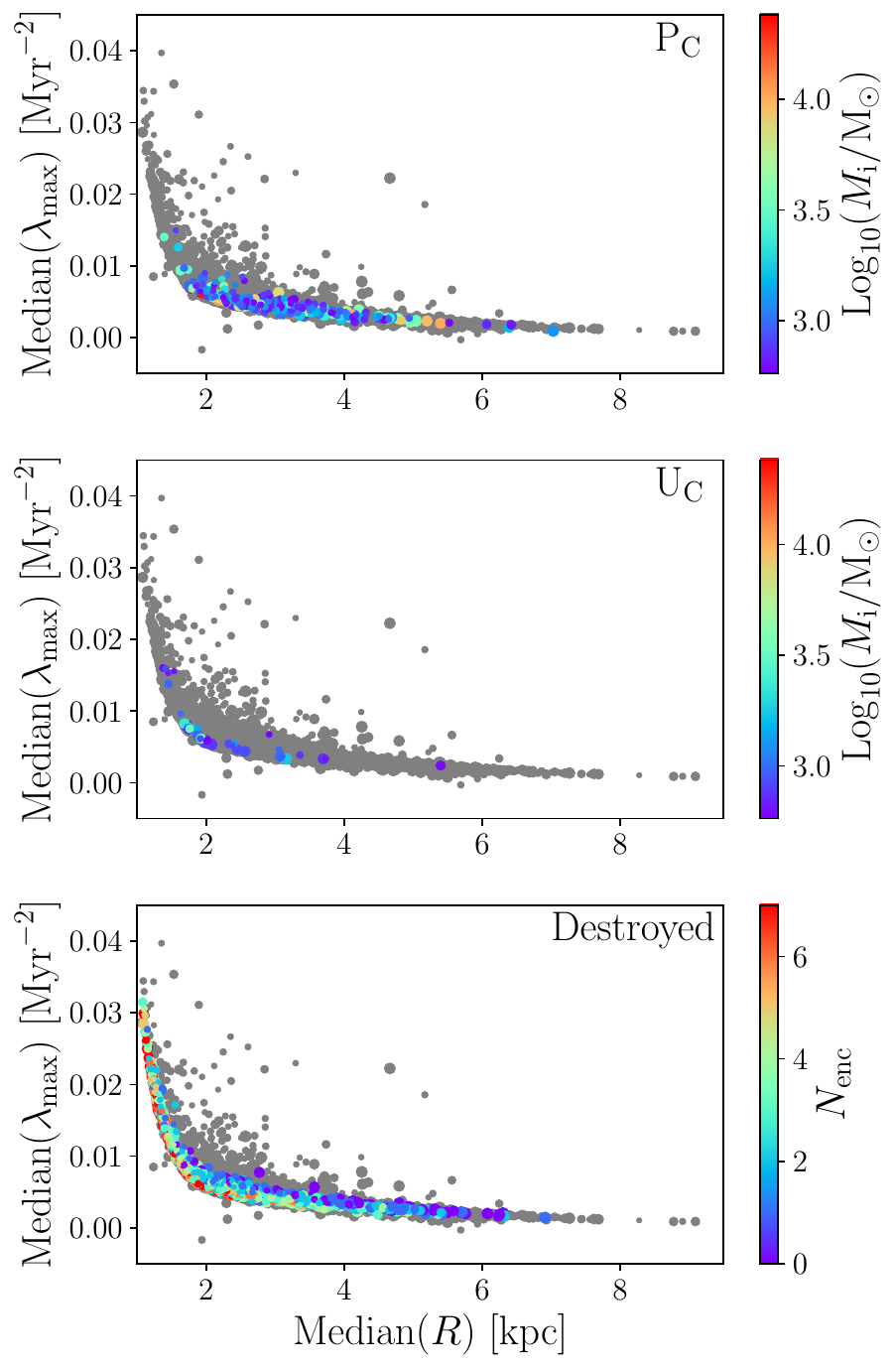}
 \caption{The median of the maximum eigenvalue from the tidal tensor as a function of median galactocentric radius for P$_{\rm{C}}$, U$_{\rm{C}}$, N$_{\rm{C}}$ and destroyed clusters. The N$_{\rm{C}}$ clusters can be seen in grey for all three plots.}
 \label{fig:EVmedian}
\end{figure}

In most cases, the majority of a cluster's lifetime is not dominated by strong GMC encounters. But rather, it is spent interacting with the rest of the galactic field. As a cluster orbits its galaxy, it will experience radial migration caused by the spiral arms \citep{SellWood2002}, while also having the possibility to be scattered above the galactic plane by GMCs \citep{Gustafsson2016, Jorgensen2020}. It is therefore important to consider the strength of the local tidal field during the cluster lifetime. To get an estimate of the strength of the tidal field that each cluster experiences, we computed the median of $\lambda_{\rm{max}}$ over the cluster lifetime. The tidal strength can be seen in Figure \ref{fig:EVmedian} as a function of the median galactocentric radius for the different cluster groups, where the grey points refer to the N$_{\rm{C}}$ clusters, and the point sizes are scaled to the relative initial half-mass radius of each cluster. Most P$_{\rm{C}}$ clusters lie between 1.5 and 6 kpc whereas the range is between 1.5 and 4 kpc for the U$_{\rm{C}}$ clusters. Even so, the tidal strength experienced is very similar. The initial cluster mass is colour coded and shows that many of the P$_{\rm{C}}$ and U$_{\rm{C}}$ clusters have low initial masses which also make them more sensitive to GMC encounters. The majority of N$_{\rm{C}}$ clusters experience a greater tidal field compared to the P$_{\rm{C}}$ and U$_{\rm{C}}$ clusters which means that GMC encounters will have less impact on N$_{\rm{C}}$ clusters compared to the rest of the galactic tidal field. This explains the smaller difference between C$_{\rm{R}}$ and C$_{\rm{N}}$ clusters. In Figure \ref{fig:EVmedian}, we also show the destroyed C$_{\rm{R}}$ clusters where the number of GMC encounters experienced in \textsc{nbody6tt} has been colour coded. The number of GMC encounters increases as the clusters spend more of their lifetime closer to the galactic centre. At a galactocentric distance of 1.3 kpc, the median tidal strength is a factor of 3 higher compared to what most of the P$_{\rm{C}}$ and U$_{\rm{C}}$ clusters experience. This increased tidal strength with the combination of an increase of GMC encounters is what ultimately destroys these clusters. For merging galaxies the tidal forces experienced by a cluster can be up to a factor of 10 to 100 stronger compared to the tides our clusters experience in M51 \citep{Li2022}. During such an event, we would expect to see an enormous increase in the disruption of clusters and probably a total destruction for most of the low-mass clusters.

\subsection{Virial state of the cluster population}
\begin{figure}
 \includegraphics[width=\columnwidth]{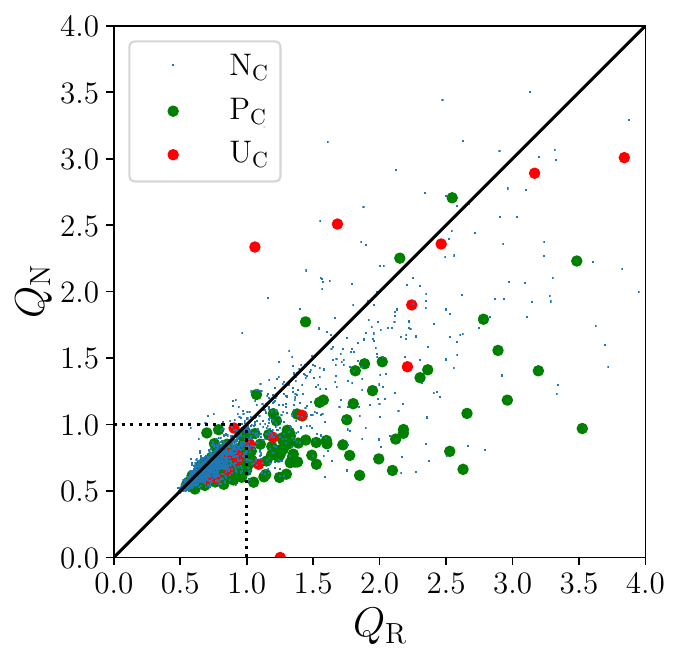}
 \caption{The virial ratio for C$_{\rm{R}}$ and C$_{\rm{N}}$ clusters. The solid line represents equality between the cluster groups, and the dotted square represents the threshold for when a cluster becomes super-virial. }
 \label{fig:Qparameter}
\end{figure}

The state of a cluster can be investigated by calculating the virial ratio:
\begin{equation}
Q = \frac{T}{|W|}
\end{equation} 
between the kintetic energy, $T$, and potential energy, $W$, of the cluster. The virial theorem tells us that a cluster will be in virial equilibrium when $Q = 0.5$. The virial ratio for C$_{\rm{R}}$ and C$_{\rm{N}}$ clusters are shown in Figure \ref{fig:Qparameter} for the different cluster groups. The solid line represents the ratio of unity between the C$_{\rm{R}}$ and C$_{\rm{N}}$ clusters and the dotted box shows the threshold for when a cluster is super-virialized. The majority of N$_{\rm{C}}$ clusters are close to being virialized, but as they become super-virialized there is strong tendency for the C$_{\rm{R}}$ clusters to have a higher virial ratio. As a cluster experiences GMC encounters, it will be subjected to a combination of extensive and compressive tides. The extensive tides will cause the cluster to expand to a point where it will be stripped of the stars in its outer parts. The compressive tides can cause the cluster to become super-virialized, as shown by \citet{Webb2017}. As the cluster moves from a compressive to an extensive tidal field, the super-virialized cluster will undergo rapid expansion and experience a significant mass loss until it reaches virial equilibrium. This seems to be the cause of most of the super-virialized clusters in all three cluster groups. The majority of P$_{\rm{C}}$ and U$_{\rm{C}}$ clusters have a virial ratio below 1.0 in the case of the C$_{\rm{N}}$ simulations. For the C$_{\rm{R}}$ simulations, 44 and 69 per cent are super-virialized for the P$_{\rm{C}}$ and U$_{\rm{C}}$ clusters, respectively. The higher fraction of super-virialized clusters for the U$_{\rm{C}}$ clusters do indeed indicate that the clusters experience more compressive tides which temporarily halts the escape of stars. This causes the C$_{\rm{R}}$ clusters to lose less mass than the C$_{\rm{N}}$ clusters, but simultaneously causes the clusters to become super-virialized. For one of the U$_{\rm{C}}$ clusters, we also see that it is destroyed in the C$_{\rm{N}}$ simulation, but survives in the C$_{\rm{R}}$ simulation.        

\begin{figure*}
 \includegraphics[width=1.0\textwidth]{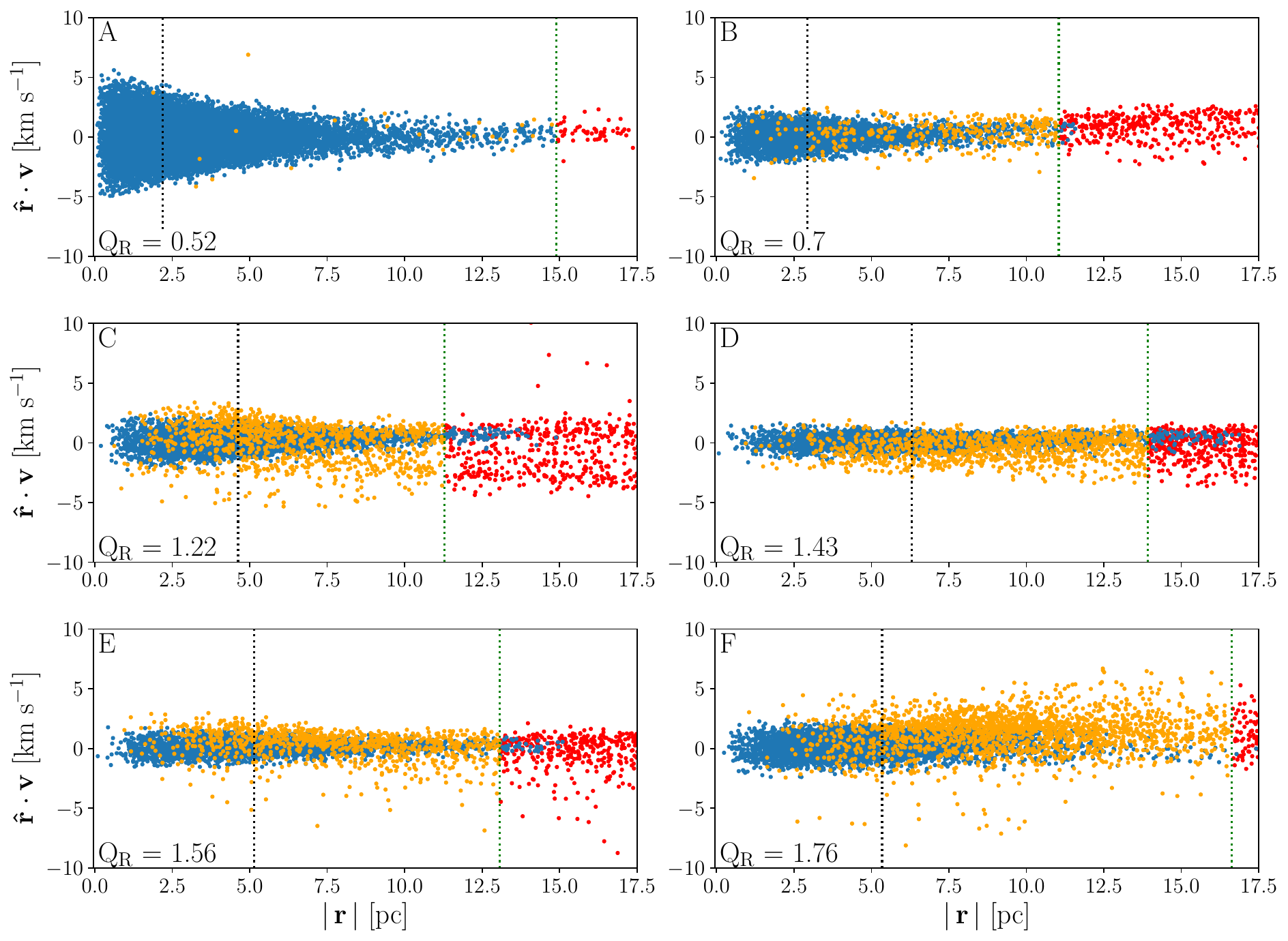}
 \caption{The dot product between the unit vector, $\hat{\boldsymbol{r}}$, and the velocity vector, $\boldsymbol{v}$, of stars as a function of the radial extent of the stars in 6 C$_{\rm{R}}$ clusters. The clusters span a range in virial ratios where cluster A represents a cluster in virial equilibrium and the following clusters show higher and higher virial ratios. The stars are divided into three different groups: \textit{blue:} bound stars, \textit{orange:} stars with positive energies that are still classified as bound because they are within the tidal radius, and \textit{red:} unbound stars.}
 \label{fig:Q6}
\end{figure*}

Figure \ref{fig:Q6} shows the dot product between the unit vector, $\hat{\boldsymbol{r}}$, and the velocity vector, $\boldsymbol{v}$, of stars as a function of the radial extent of the stars in 6 C$_{\rm{R}}$ clusters with a range of different virial ratios. The stars are colour coded into three groups: \textit{blue:} bound stars, \textit{orange:} potential escapers, i.e. stars with positive energies that are still classified as bound because they are within the tidal radius, and \textit{red:} unbound stars. $\hat{\boldsymbol{r}} \cdot \boldsymbol{v}$ indicates in which direction a star is moving, where a positive value indicate that the star is moving away from the cluster centre and a negative value indicate that the star is moving towards it. The black and green dotted vertical lines indicate the half-mass and tidal radius, respectively. Each of the clusters has been labelled from A to F. For a cluster in virial equilibrium, we would expect an equal amount of stars to be moving in both directions. This is indeed the case for cluster A and B, whereas this is only true for the bound stars in the other clusters. Clusters that have a high virial ratio also have a high fraction of potential escapers with values of 19, 27, 23 and 26 per cent for cluster C, D, E and F, respectively. This is a reflection of the chaotic and violent tidal field which the clusters have experienced. Not only are these clusters super-virialized, but their potential escapers also show a high anisotropy and have a less dense centre in terms of bound stars compared to cluster A and B. The anisotropy for the potential escapers is especially high for cluster F, where the majority of them are moving away from the cluster centre, indicating that these stars could be in the process of escaping the cluster. For the super-virialized clusters, we also see a number of bound stars that are beyond the tidal radius. \citet{Kupper2012} showed that bound stars outside the tidal radius can be recaptured by clusters. This is because if the cluster is close to its perigalacticon, the tidal radius will be much smaller and will increase at a later orbital phase. Due to the chaotic nature of the GMCs, we assume that this is also the case for our clusters where the tidal radius can be altered by the presence of a GMC, which is why we consider these stars as cluster members. Cluster D have more potential escapers which are moving towards the cluster centre, indicating that the cluster has been compressed by tides. Figure \ref{fig:Q6} reveals the difficulty of estimating cluster membership of stars and even though a cluster might have a high virial ratio, the majority of its stars can still currently be bound. Clusters that are super-virialized also display a bimodial velocity distribution which is a result of the different velocity distributions of the bound stars and the potential escapers. For clusters where their individual stars can be resolved, observing a bimodial velocity distribution could therefore indicate recent interactions with strong tidal fields.

\section{Conclusions}
\label{sec:Conclusion}
Using a galactic model of M51 combined with $N$-body simulations, we have investigated the evolution of 5000 unique clusters with ages up to 200 Myr in the mass range $[600 - 24000]$ \Mo. We have investigated the effect of GMC encounters on the clusters by performing two $N$-body simulations for each cluster: one version with the presence of GMCs, called C$_{\rm{R}}$ clusters, and another version where the tidal effects from the GMCs have been removed called C$_{\rm{N}}$ clusters.   

We find a mass function with a power-law slope of $\beta = -1.79\pm 0.02$, indicating a disruption of lower mass clusters. If a mass cut of 5000 \Mo is used, we find $\beta = -1.98\pm 0.17$ in agreement with observations of \citetalias{Messa2018}. Splitting the mass function into different age groups reveals a slope in agreement with the initial mass function slope of $-2$ for the youngest clusters. The oldest cluster group has a slope of $\beta = -1.67\pm 0.04$, indicating that most disruption happens for old low-mass clusters. 

The age function shows the most disruption for the low-mass clusters in the interval $[0.6 - 3.0] \times 10^3$ \Mo with a slope of $\gamma = -0.43\pm 0.08$ which flattens as the cluster mass interval increases. The same trend is seen for the C$_{\rm{N}}$ clusters, but with less disruption in each mass interval. We can directly compare our results for clusters in the mass interval $[0.5 - 1.0] \times 10^4$ \Mo with the observations of \citetalias{Messa2018}. We find a large discrepancy with a slope of the age function $\gamma = -0.24\pm 0.06$ compared to an observed slope of $\gamma = -0.39\pm 0.04$. We investigated if this discrepancy could be caused by the high initial velocity dispersion of 35 \kmso which the cluster and GMCs are born with in our galactic model. By comparing the mass loss of a typical GMC encounter to the mass loss of a GMC encounter with a much lower relative velocity of 10 \kmso, we were able to rescale the masses of the C$_{\rm{R}}$ clusters. We found no significant difference and conclude that the disruptiveness of the GMC encounters cannot explain the discrepancy in the age function. Instead, we propose that the discrepancy might be caused by an incompleteness in the observational data for clusters of mass $\sim$ 5000 \Mo which was also addressed by \citetalias{Messa2018}. This is further supported by the fact that we see good agreement between our results and the observations of \citetalias{Messa2018} for clusters with masses above $10^4$ \MO. 

The fraction of clusters that survive is strongly associated with the age of the clusters, especially for low-mass clusters. We find that clusters in the mass range $[0.6 - 3.0] \times 10^3$ \Mo has the least fraction of survivors after 200 Myr of $\sim$40 per cent, where a clear destruction is already seen after 50 Myr. For the C$_{\rm{N}}$ clusters, the survival fraction is $\sim$8 per cent points higher compared to the C$_{\rm{R}}$ clusters, for clusters with initial masses below $\sim$ 6000 \Mo at ages of 200 Myr.  

Investigating the environmental dependence of the mass- and age function, we divided the clusters into different radial regions, together with a spiral arm (SA) and inter-arm (IA) region. We find that the slope of the mass function is steepest with $\beta = -1.68\pm 0.05$ for the clusters that are located closest to the galactic centre. As the clusters are located further from the galactic centre, the slope of the mass function flattens. If a mass cut of 5000 \Mo is used, we find similar results as the observations of \citetalias{Messa2018b} with a slope of $\sim-2$ for all radial bins. For the SA and IA region, we find a similar trend as observations for the mass function, where the IA region has a significant steeper slope compared to the SA region. For the age function, we find most disrupting for clusters close to the galactic centre with $\gamma = -0.57\pm 0.10$. As the distance from the galactic centre increases, we see less disruption. We find good agreement with the observations of \citetalias{Messa2018b}, except for clusters closest to the galactic centre and in the SA region. The discrepancy between our results and the observations of \citetalias{Messa2018b} is likely related to an incompletness in the observations for clusters with masses of $\sim$5000 \MO.    

For the mass function, there is no clear difference between the C$_{\rm{R}}$ and C$_{\rm{N}}$ clusters, suggesting that we cannot detect the effects of the GMCs in the mass function which is related to the short age range we are investigating. The age functions have genereally less disruption for the C$_{\rm{N}}$ clusters, but still within the uncertainty of the age slope. The biggest effect of the GMCs are found by directly comparing the individual C$_{\rm{R}}$ and C$_{\rm{N}}$ clusters. We find that for the majority of clusters there is little difference between C$_{\rm{R}}$ and C$_{\rm{N}}$ clusters because these clusters do not experience any GMC encounters. We find 192 clusters with large difference, as a result of GMC encounters which have removed a significant amount of stars from the C$_{\rm{R}}$ clusters compared to the C$_{\rm{N}}$ clusters. In rare cases, we find that the presence of GMCs can actually protect the C$_{\rm{R}}$ clusters from the rest of the galactic tides which results in a significant lower amount of stars in the C$_{\rm{N}}$ clusters. 

The majority of clusters are virialized. However, for clusters that are super-virialized, we find that the majority of C$_{\rm{R}}$ clusters have a significant higher virial ratio compared to the C$_{\rm{N}}$ clusters as a result of GMC encounters. 

We are not able to detect any significant distinction between the population of the C$_{\rm{R}}$ and C$_{\rm{N}}$ clusters, and we can therefore conclude that GMCs have no detectable impact on the cluster population. This is due to the fact, that the clusters are not sufficiently old enough for the GMCs to have had a strong impact on shaping the cluster population.

%- Mass function: We observed a more flattend mass function with a slope of $-1.79\pm 0.02$ compared to observations which 

\section*{Acknowledgements}
We would like to thank Angela Adamo, Matteo Messa and Sean Linden for their valuable comments and suggestions which helped improve this paper. We would also like to thank the anonymous referee for input that has helped strengthen this paper. The $N$-body computations were enabled by resources provided by LUNARC, The Centre for Scientific and Technical Computing at Lund University, which were possible thanks to grants from The Royal Physiographic Society of Lund.
%The computations/data handling/[SIMILAR] were/was enabled by resources provided by LUNARC, The Centre for Scientific and Technical Computing at Lund University.

%Also fysiografen.....

%%%%%%%%%%%%%%%%%%%%%%%%%%%%%%%%%%%%%%%%%%%%%%%%%%
\section*{Data Availability}
The data underlying this article will be shared on reasonable request to the corresponding author.
 
%The inclusion of a Data Availability Statement is a requirement for articles published in MNRAS. Data Availability Statements provide a standardised format for readers to understand the availability of data underlying the research results described in the article. The statement may refer to original data generated in the course of the study or to third-party data analysed in the article. The statement should describe and provide means of access, where possible, by linking to the data or providing the required accession numbers for the relevant databases or DOIs.

%%%%%%%%%%%%%%%%%%%% REFERENCES %%%%%%%%%%%%%%%%%%

% The best way to enter references is to use BibTeX:

\bibliographystyle{mnras}
\bibliography{references} % if your bibtex file is called example.bib

\begin{thebibliography}{}
\makeatletter
\relax
\def\mn@urlcharsother{\let\do\@makeother \do\$\do\&\do\#\do\^\do\_\do\%\do\~}
\def\mn@doi{\begingroup\mn@urlcharsother \@ifnextchar [ {\mn@doi@}
  {\mn@doi@[]}}
\def\mn@doi@[#1]#2{\def\@tempa{#1}\ifx\@tempa\@empty \href
  {http://dx.doi.org/#2} {doi:#2}\else \href {http://dx.doi.org/#2} {#1}\fi
  \endgroup}
\def\mn@eprint#1#2{\mn@eprint@#1:#2::\@nil}
\def\mn@eprint@arXiv#1{\href {http://arxiv.org/abs/#1} {{\tt arXiv:#1}}}
\def\mn@eprint@dblp#1{\href {http://dblp.uni-trier.de/rec/bibtex/#1.xml}
  {dblp:#1}}
\def\mn@eprint@#1:#2:#3:#4\@nil{\def\@tempa {#1}\def\@tempb {#2}\def\@tempc
  {#3}\ifx \@tempc \@empty \let \@tempc \@tempb \let \@tempb \@tempa \fi \ifx
  \@tempb \@empty \def\@tempb {arXiv}\fi \@ifundefined
  {mn@eprint@\@tempb}{\@tempb:\@tempc}{\expandafter \expandafter \csname
  mn@eprint@\@tempb\endcsname \expandafter{\@tempc}}}

\bibitem[\protect\citeauthoryear{{Aarseth}}{{Aarseth}}{2003}]{Aarseth2003}
{Aarseth} S.~J.,  2003, {Gravitational N-Body Simulations}

\bibitem[\protect\citeauthoryear{Baumgardt, Hut  \& Heggie}{Baumgardt
  et~al.}{2002}]{Baumgardt2002}
Baumgardt H.,  Hut P.,   Heggie D.~C.,  2002, \mn@doi [Monthly Notices of the
  Royal Astronomical Society] {10.1046/j.1365-8711.2002.05736.x}, 336, 1069

\bibitem[\protect\citeauthoryear{{Berentzen} \& {Athanassoula}}{{Berentzen} \&
  {Athanassoula}}{2012}]{Berentzen2012}
{Berentzen} I.,  {Athanassoula} E.,  2012, \mn@doi [\mnras]
  {10.1111/j.1365-2966.2011.19964.x}, \href
  {https://ui.adsabs.harvard.edu/abs/2012MNRAS.419.3244B} {419, 3244}

\bibitem[\protect\citeauthoryear{{Bigiel}, {Leroy}, {Walter}, {Brinks}, {de
  Blok}, {Madore}  \& {Thornley}}{{Bigiel} et~al.}{2008}]{Bigiel2008}
{Bigiel} F.,  {Leroy} A.,  {Walter} F.,  {Brinks} E.,  {de Blok} W.~J.~G.,
  {Madore} B.,   {Thornley} M.~D.,  2008, \mn@doi [\aj]
  {10.1088/0004-6256/136/6/2846}, \href
  {https://ui.adsabs.harvard.edu/abs/2008AJ....136.2846B} {136, 2846}

\bibitem[\protect\citeauthoryear{Bonnell \& Davies}{Bonnell \&
  Davies}{1998}]{Bonnell1998}
Bonnell I.~A.,  Davies M.~B.,  1998, \mn@doi [Monthly Notices of the Royal
  Astronomical Society] {10.1046/j.1365-8711.1998.01372.x}, 295, 691

\bibitem[\protect\citeauthoryear{{Brown} \& {Gnedin}}{{Brown} \&
  {Gnedin}}{2021}]{Brown2021}
{Brown} G.,  {Gnedin} O.~Y.,  2021, \mn@doi [\mnras] {10.1093/mnras/stab2907},
  \href {https://ui.adsabs.harvard.edu/abs/2021MNRAS.508.5935B} {508, 5935}

\bibitem[\protect\citeauthoryear{{Chandar}, {Whitmore}, {Dinino}, {Kennicutt},
  {Chien}, {Schinnerer}  \& {Meidt}}{{Chandar} et~al.}{2016}]{Chandar2016}
{Chandar} R.,  {Whitmore} B.~C.,  {Dinino} D.,  {Kennicutt} R.~C.,  {Chien}
  L.~H.,  {Schinnerer} E.,   {Meidt} S.,  2016, \mn@doi [\apj]
  {10.3847/0004-637X/824/2/71}, \href
  {https://ui.adsabs.harvard.edu/abs/2016ApJ...824...71C} {824, 71}

\bibitem[\protect\citeauthoryear{{Chernoff} \& {Weinberg}}{{Chernoff} \&
  {Weinberg}}{1990}]{Chernoff1990}
{Chernoff} D.~F.,  {Weinberg} M.~D.,  1990, \mn@doi [\apj] {10.1086/168451},
  \href {https://ui.adsabs.harvard.edu/abs/1990ApJ...351..121C} {351, 121}

\bibitem[\protect\citeauthoryear{{Colombo} et~al.,}{{Colombo}
  et~al.}{2014a}]{Colombo2014}
{Colombo} D.,  et~al., 2014a, \mn@doi [\apj] {10.1088/0004-637X/784/1/3}, \href
  {http://adsabs.harvard.edu/abs/2014ApJ...784....3C} {784, 3}

\bibitem[\protect\citeauthoryear{{Colombo} et~al.,}{{Colombo}
  et~al.}{2014b}]{Colombo2014b}
{Colombo} D.,  et~al., 2014b, \mn@doi [\apj] {10.1088/0004-637X/784/1/4}, \href
  {https://ui.adsabs.harvard.edu/abs/2014ApJ...784....4C} {784, 4}

\bibitem[\protect\citeauthoryear{Cooper et~al.,}{Cooper
  et~al.}{2012}]{Cooper2012}
Cooper E.~M.,  et~al., 2012, \mn@doi [\apj] {10.1088/0004-637X/755/2/165}, 755,
  165

\bibitem[\protect\citeauthoryear{{Dobbs}, {Theis}, {Pringle}  \&
  {Bate}}{{Dobbs} et~al.}{2010}]{Dobbs2010}
{Dobbs} C.~L.,  {Theis} C.,  {Pringle} J.~E.,   {Bate} M.~R.,  2010, \mn@doi
  [\mnras] {10.1111/j.1365-2966.2009.16161.x}, \href
  {http://adsabs.harvard.edu/abs/2010MNRAS.403..625D} {403, 625}

\bibitem[\protect\citeauthoryear{{Gieles}}{{Gieles}}{2009}]{Gieles2009}
{Gieles} M.,  2009, \mn@doi [\mnras] {10.1111/j.1365-2966.2009.14473.x}, \href
  {https://ui.adsabs.harvard.edu/abs/2009MNRAS.394.2113G} {394, 2113}

\bibitem[\protect\citeauthoryear{Gieles \& Baumgardt}{Gieles \&
  Baumgardt}{2008}]{Gieles2008}
Gieles M.,  Baumgardt H.,  2008, \mn@doi [Monthly Notices of the Royal
  Astronomical Society: Letters] {10.1111/j.1745-3933.2008.00515.x}, 389, L28

\bibitem[\protect\citeauthoryear{{Gieles}, {Portegies Zwart}, {Baumgardt},
  {Athanassoula}, {Lamers}, {Sipior}  \& {Leenaarts}}{{Gieles}
  et~al.}{2006}]{Gieles2006}
{Gieles} M.,  {Portegies Zwart} S.~F.,  {Baumgardt} H.,  {Athanassoula} E.,
  {Lamers} H.~J.~G.~L.~M.,  {Sipior} M.,   {Leenaarts} J.,  2006, \mn@doi
  [\mnras] {10.1111/j.1365-2966.2006.10711.x}, \href
  {https://ui.adsabs.harvard.edu/abs/2006MNRAS.371..793G} {371, 793}

\bibitem[\protect\citeauthoryear{Gieles, Athanassoula  \&
  Portegies~Zwart}{Gieles et~al.}{2007}]{Gieles2007}
Gieles M.,  Athanassoula E.,   Portegies~Zwart S.~F.,  2007, \mn@doi [Monthly
  Notices of the Royal Astronomical Society]
  {10.1111/j.1365-2966.2007.11477.x}, 376, 809

\bibitem[\protect\citeauthoryear{Giersz \& Heggie}{Giersz \&
  Heggie}{1994}]{Giersz1994}
Giersz M.,  Heggie D.~C.,  1994, \mn@doi [Monthly Notices of the Royal
  Astronomical Society] {10.1093/mnras/268.1.257}, 268, 257

\bibitem[\protect\citeauthoryear{{Gratton}, {Bragaglia}, {Carretta}, {D'Orazi},
  {Lucatello}  \& {Sollima}}{{Gratton} et~al.}{2019}]{Gratton2019}
{Gratton} R.,  {Bragaglia} A.,  {Carretta} E.,  {D'Orazi} V.,  {Lucatello} S.,
   {Sollima} A.,  2019, \mn@doi [\aapr] {10.1007/s00159-019-0119-3}, \href
  {https://ui.adsabs.harvard.edu/abs/2019A&ARv..27....8G} {27, 8}

\bibitem[\protect\citeauthoryear{{Gustafsson}, {Church}, {Davies}  \&
  {Rickman}}{{Gustafsson} et~al.}{2016}]{Gustafsson2016}
{Gustafsson} B.,  {Church} R.~P.,  {Davies} M.~B.,   {Rickman} H.,  2016,
  \mn@doi [\aap] {10.1051/0004-6361/201423916}, \href
  {http://adsabs.harvard.edu/abs/2016A%26A...593A..85G} {593, A85}

\bibitem[\protect\citeauthoryear{{Hagiwara}}{{Hagiwara}}{2007}]{Hagiwara2007}
{Hagiwara} Y.,  2007, \mn@doi [\aj] {10.1086/510383}, \href
  {https://ui.adsabs.harvard.edu/abs/2007AJ....133.1176H} {133, 1176}

\bibitem[\protect\citeauthoryear{{Hernquist}}{{Hernquist}}{1990}]{Hernquist1990}
{Hernquist} L.,  1990, \mn@doi [\apj] {10.1086/168845}, \href
  {https://ui.adsabs.harvard.edu/abs/1990ApJ...356..359H} {356, 359}

\bibitem[\protect\citeauthoryear{{Holmberg}, {Nordstr{\"o}m}  \&
  {Andersen}}{{Holmberg} et~al.}{2009}]{Holmberg2009}
{Holmberg} J.,  {Nordstr{\"o}m} B.,   {Andersen} J.,  2009, \mn@doi [\aap]
  {10.1051/0004-6361/200811191}, \href
  {http://adsabs.harvard.edu/abs/2009A%26A...501..941H} {501, 941}

\bibitem[\protect\citeauthoryear{{Hopkins}, {Quataert}  \& {Murray}}{{Hopkins}
  et~al.}{2012}]{Hopkins2012}
{Hopkins} P.~F.,  {Quataert} E.,   {Murray} N.,  2012, \mn@doi [\mnras]
  {10.1111/j.1365-2966.2012.20578.x}, \href
  {https://ui.adsabs.harvard.edu/abs/2012MNRAS.421.3488H} {421, 3488}

\bibitem[\protect\citeauthoryear{{Hurley} \& {Mackey}}{{Hurley} \&
  {Mackey}}{2010}]{Hurley2010}
{Hurley} J.~R.,  {Mackey} A.~D.,  2010, \mn@doi [\mnras]
  {10.1111/j.1365-2966.2010.17285.x}, \href
  {https://ui.adsabs.harvard.edu/abs/2010MNRAS.408.2353H} {408, 2353}

\bibitem[\protect\citeauthoryear{{J{\o}rgensen} \& {Church}}{{J{\o}rgensen} \&
  {Church}}{2020}]{Jorgensen2020}
{J{\o}rgensen} T.~G.,  {Church} R.~P.,  2020, \mn@doi [\mnras]
  {10.1093/mnras/staa185}, \href
  {https://ui.adsabs.harvard.edu/abs/2020MNRAS.492.4959J} {492, 4959}

\bibitem[\protect\citeauthoryear{Koda et~al.,}{Koda et~al.}{2009}]{Koda2009}
Koda J.,  et~al., 2009, \mn@doi [The Astrophysical Journal]
  {10.1088/0004-637X/700/2/L132}, 700, L132

\bibitem[\protect\citeauthoryear{{Kroupa}}{{Kroupa}}{2001}]{Kroupa2001}
{Kroupa} P.,  2001, \mn@doi [\mnras] {10.1046/j.1365-8711.2001.04022.x}, \href
  {http://adsabs.harvard.edu/abs/2001MNRAS.322..231K} {322, 231}

\bibitem[\protect\citeauthoryear{{K{\"u}pper}, {Lane}  \&
  {Heggie}}{{K{\"u}pper} et~al.}{2012}]{Kupper2012}
{K{\"u}pper} A. H.~W.,  {Lane} R.~R.,   {Heggie} D.~C.,  2012, \mn@doi [\mnras]
  {10.1111/j.1365-2966.2011.20242.x}, \href
  {https://ui.adsabs.harvard.edu/abs/2012MNRAS.420.2700K} {420, 2700}

\bibitem[\protect\citeauthoryear{Lamers, Baumgardt  \& Gieles}{Lamers
  et~al.}{2010}]{Lamers2010}
Lamers H. J. G. L.~M.,  Baumgardt H.,   Gieles M.,  2010, \mn@doi [Monthly
  Notices of the Royal Astronomical Society]
  {10.1111/j.1365-2966.2010.17309.x}, 409, 305

\bibitem[\protect\citeauthoryear{{Li}, {Vogelsberger}, {Bryan}, {Marinacci},
  {Sales}  \& {Torrey}}{{Li} et~al.}{2022}]{Li2022}
{Li} H.,  {Vogelsberger} M.,  {Bryan} G.~L.,  {Marinacci} F.,  {Sales} L.~V.,
  {Torrey} P.,  2022, \mn@doi [\mnras] {10.1093/mnras/stac1136}, \href
  {https://ui.adsabs.harvard.edu/abs/2022MNRAS.514..265L} {514, 265}

\bibitem[\protect\citeauthoryear{{Linden} et~al.,}{{Linden}
  et~al.}{2022}]{Linden2022}
{Linden} S.~T.,  et~al., 2022, \mn@doi [\apj] {10.3847/1538-4357/ac7c07}, \href
  {https://ui.adsabs.harvard.edu/abs/2022ApJ...935..166L} {935, 166}

\bibitem[\protect\citeauthoryear{{Lynden-Bell}}{{Lynden-Bell}}{1999}]{LyndenBell1999}
{Lynden-Bell} D.,  1999, \mn@doi [Physica A Statistical Mechanics and its
  Applications] {10.1016/S0378-4371(98)00518-4}, \href
  {https://ui.adsabs.harvard.edu/abs/1999PhyA..263..293L} {263, 293}

\bibitem[\protect\citeauthoryear{{Lynden-Bell} \& {Wood}}{{Lynden-Bell} \&
  {Wood}}{1968}]{LyndenBell1968}
{Lynden-Bell} D.,  {Wood} R.,  1968, \mn@doi [\mnras]
  {10.1093/mnras/138.4.495}, \href
  {https://ui.adsabs.harvard.edu/abs/1968MNRAS.138..495L} {138, 495}

\bibitem[\protect\citeauthoryear{{Martinez-Medina}, {Pichardo}, {Peimbert}  \&
  {Moreno}}{{Martinez-Medina} et~al.}{2017}]{Martinez2017}
{Martinez-Medina} L.~A.,  {Pichardo} B.,  {Peimbert} A.,   {Moreno} E.,  2017,
  \mn@doi [\apj] {10.3847/1538-4357/834/1/58}, \href
  {https://ui.adsabs.harvard.edu/abs/2017ApJ...834...58M} {834, 58}

\bibitem[\protect\citeauthoryear{{McQuinn}, {Skillman}, {Dolphin}, {Berg}  \&
  {Kennicutt}}{{McQuinn} et~al.}{2016}]{McQuinn2016}
{McQuinn} K. B.~W.,  {Skillman} E.~D.,  {Dolphin} A.~E.,  {Berg} D.,
  {Kennicutt} R.,  2016, \mn@doi [\apj] {10.3847/0004-637X/826/1/21}, \href
  {https://ui.adsabs.harvard.edu/abs/2016ApJ...826...21M} {826, 21}

\bibitem[\protect\citeauthoryear{{Meidt}, {Rand}, {Merrifield}, {Shetty}  \&
  {Vogel}}{{Meidt} et~al.}{2008}]{Meidt2008}
{Meidt} S.~E.,  {Rand} R.~J.,  {Merrifield} M.~R.,  {Shetty} R.,   {Vogel}
  S.~N.,  2008, \mn@doi [\apj] {10.1086/591516}, \href
  {https://ui.adsabs.harvard.edu/abs/2008ApJ...688..224M} {688, 224}

\bibitem[\protect\citeauthoryear{{Meidt} et~al.,}{{Meidt}
  et~al.}{2015}]{Meidt2015}
{Meidt} S.~E.,  et~al., 2015, \mn@doi [\apj] {10.1088/0004-637X/806/1/72},
  \href {http://adsabs.harvard.edu/abs/2015ApJ...806...72M} {806, 72}

\bibitem[\protect\citeauthoryear{{Messa} et~al.,}{{Messa}
  et~al.}{2018a}]{Messa2018}
{Messa} M.,  et~al., 2018a, \mn@doi [\mnras] {10.1093/mnras/stx2403}, \href
  {http://adsabs.harvard.edu/abs/2018MNRAS.473..996M} {473, 996}

\bibitem[\protect\citeauthoryear{{Messa} et~al.,}{{Messa}
  et~al.}{2018b}]{Messa2018b}
{Messa} M.,  et~al., 2018b, \mn@doi [\mnras] {10.1093/mnras/sty577}, \href
  {http://adsabs.harvard.edu/abs/2018MNRAS.477.1683M} {477, 1683}

\bibitem[\protect\citeauthoryear{{Oikawa} \& {Sofue}}{{Oikawa} \&
  {Sofue}}{2014}]{Oikawa2014}
{Oikawa} S.,  {Sofue} Y.,  2014, \mn@doi [\pasj] {10.1093/pasj/psu059}, \href
  {https://ui.adsabs.harvard.edu/abs/2014PASJ...66...77O} {66, 77}

\bibitem[\protect\citeauthoryear{{Oort}}{{Oort}}{1958}]{Oort1958}
{Oort} J.~H.,  1958, Ricerche Astronomiche, \href
  {https://ui.adsabs.harvard.edu/abs/1958RA......5..507O} {5, 507}

\bibitem[\protect\citeauthoryear{{Ostriker}, {Spitzer}  \&
  {Chevalier}}{{Ostriker} et~al.}{1972}]{Ostriker1972}
{Ostriker} J.~P.,  {Spitzer} Jr. L.,   {Chevalier} R.~A.,  1972, \mn@doi
  [\apjl] {10.1086/181018}, \href
  {https://ui.adsabs.harvard.edu/abs/1972ApJ...176L..51O} {176, L51}

\bibitem[\protect\citeauthoryear{{Pety} et~al.,}{{Pety}
  et~al.}{2013}]{Pety2013}
{Pety} J.,  et~al., 2013, \mn@doi [\apj] {10.1088/0004-637X/779/1/43}, \href
  {https://ui.adsabs.harvard.edu/abs/2013ApJ...779...43P} {779, 43}

\bibitem[\protect\citeauthoryear{{Pichardo}, {Martos}, {Moreno}  \&
  {Espresate}}{{Pichardo} et~al.}{2003}]{Pichardo2003}
{Pichardo} B.,  {Martos} M.,  {Moreno} E.,   {Espresate} J.,  2003, \mn@doi
  [\apj] {10.1086/344592}, \href
  {http://adsabs.harvard.edu/abs/2003ApJ...582..230P} {582, 230}

\bibitem[\protect\citeauthoryear{{Plummer}}{{Plummer}}{1911}]{Plummer1911}
{Plummer} H.~C.,  1911, \mn@doi [\mnras] {10.1093/mnras/71.5.460}, \href
  {https://ui.adsabs.harvard.edu/abs/1911MNRAS..71..460P} {71, 460}

\bibitem[\protect\citeauthoryear{{Renaud}}{{Renaud}}{2018}]{Renaud2018}
{Renaud} F.,  2018, \mn@doi [\nar] {10.1016/j.newar.2018.03.001}, \href
  {https://ui.adsabs.harvard.edu/abs/2018NewAR..81....1R} {81, 1}

\bibitem[\protect\citeauthoryear{{Renaud}, {Boily}, {Fleck}, {Naab}  \&
  {Theis}}{{Renaud} et~al.}{2008}]{Renaud2008}
{Renaud} F.,  {Boily} C.~M.,  {Fleck} J.~J.,  {Naab} T.,   {Theis} C.,  2008,
  \mn@doi [\mnras] {10.1111/j.1745-3933.2008.00564.x}, \href
  {https://ui.adsabs.harvard.edu/abs/2008MNRAS.391L..98R} {391, L98}

\bibitem[\protect\citeauthoryear{{Renaud}, {Boily}, {Naab}  \&
  {Theis}}{{Renaud} et~al.}{2009}]{Renaud2009}
{Renaud} F.,  {Boily} C.~M.,  {Naab} T.,   {Theis} C.,  2009, \mn@doi [\apj]
  {10.1088/0004-637X/706/1/67}, \href
  {https://ui.adsabs.harvard.edu/abs/2009ApJ...706...67R} {706, 67}

\bibitem[\protect\citeauthoryear{{Renaud}, {Gieles}  \& {Boily}}{{Renaud}
  et~al.}{2011}]{Renaud2011}
{Renaud} F.,  {Gieles} M.,   {Boily} C.~M.,  2011, \mn@doi [\mnras]
  {10.1111/j.1365-2966.2011.19531.x}, \href
  {http://adsabs.harvard.edu/abs/2011MNRAS.418..759R} {418, 759}

\bibitem[\protect\citeauthoryear{{Salo} \& {Laurikainen}}{{Salo} \&
  {Laurikainen}}{2000}]{Salo2000b}
{Salo} H.,  {Laurikainen} E.,  2000, \mn@doi [\mnras]
  {10.1046/j.1365-8711.2000.03650.x}, \href
  {https://ui.adsabs.harvard.edu/abs/2000MNRAS.319..377S} {319, 377}

\bibitem[\protect\citeauthoryear{Schinnerer et~al.,}{Schinnerer
  et~al.}{2013}]{Schinnerer2013}
Schinnerer E.,  et~al., 2013, \mn@doi [The Astrophysical Journal]
  {10.1088/0004-637X/779/1/42}, 779, 42

\bibitem[\protect\citeauthoryear{{Schruba} et~al.,}{{Schruba}
  et~al.}{2011}]{Schruba2011}
{Schruba} A.,  et~al., 2011, \mn@doi [\aj] {10.1088/0004-6256/142/2/37}, \href
  {https://ui.adsabs.harvard.edu/abs/2011AJ....142...37S} {142, 37}

\bibitem[\protect\citeauthoryear{{Sellwood} \& {Binney}}{{Sellwood} \&
  {Binney}}{2002}]{SellWood2002}
{Sellwood} J.~A.,  {Binney} J.~J.,  2002, \mn@doi [\mnras]
  {10.1046/j.1365-8711.2002.05806.x}, \href
  {https://ui.adsabs.harvard.edu/abs/2002MNRAS.336..785S} {336, 785}

\bibitem[\protect\citeauthoryear{Shetty, Vogel, Ostriker  \& Teuben}{Shetty
  et~al.}{2007}]{Shetty2007}
Shetty R.,  Vogel S.~N.,  Ostriker E.~C.,   Teuben P.~J.,  2007, \mn@doi [The
  Astrophysical Journal] {10.1086/520037}, 665, 1138

\bibitem[\protect\citeauthoryear{{Sofue}, {Tutui}, {Honma}, {Tomita},
  {Takamiya}, {Koda}  \& {Takeda}}{{Sofue} et~al.}{1999}]{Sofue1999}
{Sofue} Y.,  {Tutui} Y.,  {Honma} M.,  {Tomita} A.,  {Takamiya} T.,  {Koda} J.,
    {Takeda} Y.,  1999, \mn@doi [\apj] {10.1086/307731}, \href
  {http://adsabs.harvard.edu/abs/1999ApJ...523..136S} {523, 136}

\bibitem[\protect\citeauthoryear{{Spitzer}}{{Spitzer}}{1958}]{Spitzer1958}
{Spitzer} Jr. L.,  1958, \mn@doi [\apj] {10.1086/146435}, \href
  {https://ui.adsabs.harvard.edu/abs/1958ApJ...127...17S} {127, 17}

\bibitem[\protect\citeauthoryear{{Tress}, {Smith}, {Sormani}, {Glover},
  {Klessen}, {Mac Low}  \& {Clark}}{{Tress} et~al.}{2020}]{Tress2020}
{Tress} R.~G.,  {Smith} R.~J.,  {Sormani} M.~C.,  {Glover} S. C.~O.,  {Klessen}
  R.~S.,  {Mac Low} M.-M.,   {Clark} P.~C.,  2020, \mn@doi [\mnras]
  {10.1093/mnras/stz3600}, \href
  {https://ui.adsabs.harvard.edu/abs/2020MNRAS.492.2973T} {492, 2973}

\bibitem[\protect\citeauthoryear{{Tully}}{{Tully}}{1974}]{Tully1974}
{Tully} R.~B.,  1974, \mn@doi [\apjs] {10.1086/190305}, \href
  {https://ui.adsabs.harvard.edu/abs/1974ApJS...27..437T} {27, 437}

\bibitem[\protect\citeauthoryear{{Wang} et~al.,}{{Wang}
  et~al.}{2016}]{Wang2016}
{Wang} L.,  et~al., 2016, \mn@doi [\mnras] {10.1093/mnras/stw274}, \href
  {https://ui.adsabs.harvard.edu/abs/2016MNRAS.458.1450W} {458, 1450}

\bibitem[\protect\citeauthoryear{{Webb}, {Patel}  \& {Vesperini}}{{Webb}
  et~al.}{2017}]{Webb2017}
{Webb} J.~J.,  {Patel} S.~S.,   {Vesperini} E.,  2017, \mn@doi [\mnras]
  {10.1093/mnrasl/slx030}, \href
  {https://ui.adsabs.harvard.edu/abs/2017MNRAS.468L..92W} {468, L92}

\bibitem[\protect\citeauthoryear{White}{White}{1977}]{White1977}
White S. D.~M.,  1977, \mn@doi [Monthly Notices of the Royal Astronomical
  Society] {10.1093/mnras/179.2.33}, 179, 33

\bibitem[\protect\citeauthoryear{{Whitmore} et~al.,}{{Whitmore}
  et~al.}{2023}]{Whitmore2023}
{Whitmore} B.~C.,  et~al., 2023, \mn@doi [\mnras] {10.1093/mnras/stad098},
  \href {https://ui.adsabs.harvard.edu/abs/2023MNRAS.520...63W} {520, 63}

\bibitem[\protect\citeauthoryear{{Wielen}}{{Wielen}}{1971}]{Wielen1971}
{Wielen} R.,  1971, \aap, \href
  {https://ui.adsabs.harvard.edu/abs/1971A&A....13..309W} {13, 309}

\bibitem[\protect\citeauthoryear{{Williams} \& {McKee}}{{Williams} \&
  {McKee}}{1997}]{Williams1997}
{Williams} J.~P.,  {McKee} C.~F.,  1997, \mn@doi [\apj] {10.1086/303588}, \href
  {https://ui.adsabs.harvard.edu/abs/1997ApJ...476..166W} {476, 166}

\bibitem[\protect\citeauthoryear{{de Blok}, {Walter}, {Brinks}, {Trachternach},
  {Oh}  \& {Kennicutt}}{{de Blok} et~al.}{2008}]{deBlok2008}
{de Blok} W.~J.~G.,  {Walter} F.,  {Brinks} E.,  {Trachternach} C.,  {Oh}
  S.~H.,   {Kennicutt} Jr. R.~C.,  2008, \mn@doi [\aj]
  {10.1088/0004-6256/136/6/2648}, \href
  {https://ui.adsabs.harvard.edu/abs/2008AJ....136.2648D} {136, 2648}

\makeatother
\end{thebibliography}

% Alternatively you could enter them by hand, like this:
% This method is tedious and prone to error if you have lots of references
%\begin{thebibliography}{99}
%\bibitem[\protect\citeauthoryear{Author}{2012}]{Author2012}
%Author A.~N., 2013, Journal of Improbable Astronomy, 1, 1
%\bibitem[\protect\citeauthoryear{Others}{2013}]{Others2013}
%Others S., 2012, Journal of Interesting Stuff, 17, 198
%\end{thebibliography}

%%%%%%%%%%%%%%%%%%%%%%%%%%%%%%%%%%%%%%%%%%%%%%%

% Don't change these lines
\bsp	% typesetting comment
\label{lastpage}
\end{document}